\documentclass{article}

\usepackage{fullpage}
\usepackage{appendix}
\usepackage{graphicx}
\usepackage{amsmath}
\usepackage{amssymb}
\usepackage{amsthm, amsfonts}
\usepackage{algorithm, algorithmic}
\usepackage{subfig}
\usepackage{hyperref}
\usepackage{float}
\usepackage{verbatim}

\newcommand{\R}{\mathbb{R}}
\newcommand{\Z}{\mathbb{Z}}
\newcommand{\N}{\mathbb{N}}

\newcommand{\Expect}{\mathsf{E}}
\newcommand{\Var}{\mathsf{Var}}
\newcommand{\Prob}{\mathsf{Pr}}

\newcommand{\stakedist}{\pi_{\text{stake}}}
\newcommand{\lenddist}{\pi_{\text{lend}}}
\newcommand{\tokendist}{\pi_{\text{token}}}
\newcommand{\borrowdist}{\pi_{\text{borrow}}}

\newcommand{\estakedist}{\hat{\pi}_{\text{stake}}}
\newcommand{\elenddist}{\hat{\pi}_{\text{lend}}}

\newcommand{\stakeprob}{\hat{\pi}_{\text{stake}}}

\newcommand{\dstake}{\Delta^{\text{stake}}_{i}}
\newcommand{\dlend}{\Delta_{\text{lend}}}

\newcommand{\mustake}{\mu_{\text{stake}}}
\newcommand{\mulend}{\mu_{\text{lend}}}

\newcommand{\slashprob}{p_{\text{slash}}}

\newcommand{\staketime}{\tau_{\text{stake}}}
\newcommand{\lendtime}{\tau_{\text{lend}}}

\DeclareMathOperator*{\argmin}{arg\,min}
\DeclareMathOperator*{\argmax}{arg\,max}
\usepackage[numbib]{tocbibind}
\newtheorem{claim}{Claim}

\newtheorem{assumption}{Assumption}

\begin{document}
\title{Competitive equilibria between staking and on-chain lending}
\author{Tarun Chitra\footnote{Gauntlet Networks, Inc.}}
\maketitle

\begin{abstract}
Proof of Stake (PoS) is a burgeoning Sybil resistance mechanism that aims to have a digital asset (``token") serve as security collateral in crypto networks.
However, PoS has so far eluded a comprehensive threat model that encompasses both Byzantine attacks from distributed systems and financial attacks that arise from the dual usage of the token as a means of payment and a Sybil resistance mechanism.
In particular, the existence of derivatives markets makes malicious coordination among validators easier to execute than in Proof of Work systems.
We demonstrate that it is also possible for on-chain lending smart contracts to cannibalize network security in PoS systems.
When the yield provided by these contracts is more attractive than the inflation rate provided from staking, stakers will tend to remove their staked tokens and lend them out, thus reducing network security.
In this paper, we provide a simple stochastic model that describes how rational validators with varying risk preferences react to changes in staking and lending returns.
For a particular configuration of this model, we provide a formal proof of a phase transition between equilibria in which tokens are predominantly staked and those in which they are predominantly lent.
We further validate this emergent adversarial behavior (e.g. reduced staked token supply) with agent-based simulations that sample transitions under more realistic conditions.
Our results illustrate that rational, non-adversarial actors can dramatically reduce PoS network security if block rewards are not calibrated appropriately above the expected yields of on-chain lending.
\end{abstract}

\section{Introduction}
There is currently an intense effort to improve the scalability of blockchains and other decentralized value systems known as crypto networks.
These networks use cryptographic proofs and game theoretic constructions to provide tamper-resistant updates to a global ledger.
While there are a variety of research and engineering challenges in setting up these systems, one of the major bottlenecks to network throughput is the cost of Sybil resistance mechanisms within a decentralized consensus protocol.
Proof of Work (PoW) networks achieve Sybil resistance by requiring consensus-participating nodes to provably burn energy to compute many iterations of a particular cryptographic hash function.
PoW, while effective and permissionless, expends a large amount of natural resources and has resulted in concentrated ownership of the underlying digital assets (e.g. Bitcoin).
Proof of Stake (PoS) was first introduced as an alternative in a 2012 BitcoinTalk post \cite{bitcointalk2011} that showed the equivalence between a PoW miner who could immediately reinvest her block rewards into hash power within the network and a PoS validator who can reinvest their validation earnings into network security.
PoS works by instead allowing users to `lock' a digital asset, known as a token, into a smart contract that provides them with token-denominated returns in exchange for validating transactions and providing network security.
Using shared, verifiable randomness, all network participants can use a multi-party computation protocol to sample the distribution of asset ownership locked into the contract and choose participant(s) who receive the block reward emitted by the network.
This is analogous to how PoW can be thought of as a protocol that samples the distribution of hash power to choose the next block producer \cite{pass2017analysis}.
One of the main benefits of PoS is that one does not have to commit a costly natural resource to participate in the network.
Instead, a purely digital asset is used as collateral for the network and the network can control its supply to provide the desired properties.
For an introductory background on PoS protocols and their complex security models, please see \cite{kiayias2017ouroboros,king2012ppcoin,saleh2019blockchain,daian2019snow,neuder2019selfish}.

In this paper, we show that these purported benefits do not come for free.
As PoS algorithms inherently connect a decentralized network's security with the capital cost of a digital asset, PoS protocols tie their security to the cost of capital rather than to the cost of a natural resource.
Volatility in the cost of capital, which is usually higher than that of natural resources \cite{poelhekke2007volatility}, can have adverse effects on capital commitments to PoS networks.
The main result we show is that alternative sources of yield can drive staking token capital allocators to collectively drain a network's security, akin to a bank run.
In particular, we find that PoS in deflationary systems is unstable and unlikely to work and that for more reasonable inflation rates, the effectiveness of PoS depends on the relationship between staking and lending rates. 
This relationship should inform the further design of PoS systems, especially as a large number of networks are launching in 2020.

\subsection*{Transitioning Security from PoW to PoS}
The move from PoW to PoS presents a plethora of challenges.
In a PoS system, the network relies on participants who are staked in the system to stay online in order to achieve liveness.
In practice, this is implemented by slashing participants --- redistributing or burning a participant's stake that is committed for validation rewards when they perform a malicious act --- who go offline or miss a block that they are supposed to produce.
Moreover, there are attacks that are unique to PoS such as the \emph{nothing-at-stake} and \emph{long-range} attacks \cite{houy2014will,gavzi2018stake}.
These attacks are impossible in PoW, as resource costs of digital assets are practically zero, especially when compared to costs of natural resources \cite{brown2019formal}.
Lastly, as the asset used for staking is also the medium of exchange, a malicious validator needs to only aggregate 33\% of the token to perform a Byzantine attack.\footnote{The majority of Proof of Stake systems use traditional Byzantine Fault Tolerant algorithms \cite{lamport1982byzantine,castro1999practical} once validators are sampled, which is why they are only resistant to 33\% Byzantine actors versus 50\% for Nakamoto consensus} 

However, if there exist physically settled futures contracts on PoS tokens, then it is possible for an attacker to buy futures that allow for staking participants to sell their staked token in the future.
This attacker can aggregate this stake and upon reaching an attack threshold, begin to perform a double spend or other malicious attack \cite{chitra_chiang_morrow_sood_2019}.
As these derivatives can be settled off-chain (e.g. using a centralized exchange like BitMEX or Deribit), monitoring of this type of attack can be difficult.
In PoW, one would need to aggregate the hash power needed to produce 50\% of the network's hashrate, which is a much harder task that relies on aggregating data centers, specialized hardware, cheap electricity, and a favorable country of residence.
Moreover, PoS systems are vulnerable to \emph{financial attacks}, or attacks that utilize the fact that PoS tokes serve as both the instrument of security and as the medium of exchange.
Such attacks are often feasible due to emergent and unexpected coordination between participants in a PoS network and an external market.

Given the vulnerability of PoS to cartel-like behavior that can be coordinated via an external market, one might naturally ask if there are also any endogenous financial risks on PoS protocols that support smart contracts.
Recently, there has been an uptick in interest in Decentralized Finance (`DeFi'), which uses smart contracts to implement standard financial primitives in a purely on-chain manner \cite{klages2019stability}.
These primitives, such as exchanges \cite{warren20170x}, lending \cite{compound2019}, and stable reserve currencies \cite{kamvarcelo,al2017basis,team2017dai}, decentralize banking functions by creating incentives that encourage rational participants to receive arbitrage profits for maintaining the system's security, while also meting out financial punishments for misbehavior.
Instead of explicitly punishing fraud via legal recourse, these protocols use purely financial modes of recourse to encourage network participation and growth.
One of the biggest sectors within DeFi is the on-chain lending market, in which the largest single platform is Compound \cite{compound2019}, an Ethereum smart contract that allows users to lend and borrow assets that conform to the ERC-20 token standard.
The Compound smart contract has held up to \$175 million of assets, has had over 40\% of the float of the Dai stablecoin \cite{team2017dai}, and saw double digit asset growth during 2019.
Given that Ethereum is likely to transition to PoS soon one must evaluate: are there any financial attacks against chain security that result from an on-chain lending system?
A simple gedanken experiment for answering this question from the view of on-chain lending  might be of the form: \begin{quote} Suppose that we assume that validators are rational financial agents. Would they not simply move their assets between staking and on-chain lending, depending on which has a higher yield?
\end{quote} 
In particular, it is clear that there is a relationship between the price of capital availability and participant's willingness to stake, as stakers have to earn more than a risk-adjusted market rate on their staked capital.
However, unlike Proof of Work, there are no physical limits that prevent validators in staking networks from rapidly moving their assets into higher yielding activities.
On-chain lending, such as Compound, makes this particularly efficient, as validators simply have to post a single Ethereum transaction to unbond their tokens and begin earning yield within a single block time.
High lending yields would likely lead to a reduction in network security (e.g. a financial attack) as these yields would encourage rational actors to unexpectedly coordinate and reduce network security by optimizing for financial gain.

\begin{figure}
    \centering
    \includegraphics[scale=0.25]{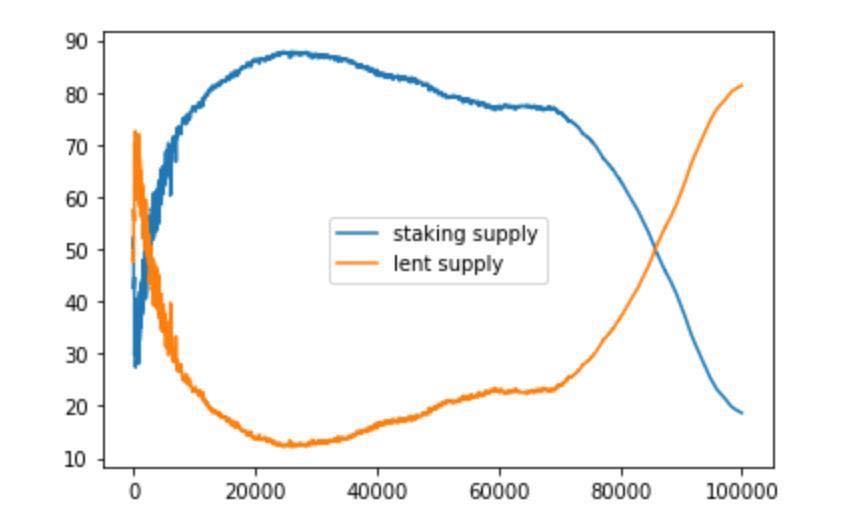}
    \includegraphics[scale=0.25]{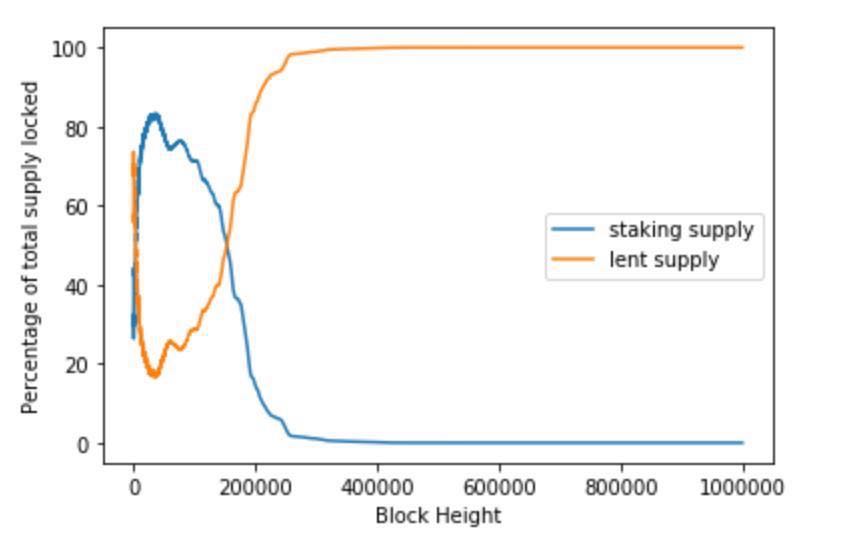}
    \caption{Staking and lending with Compound's smart contract parameters (left) and more aggressive parameters with yields of up to 75\%. The y-axis in both figures is the percentage of the total outstanding token supply that is allocated to staking versus that which is allocated to lending. This simulation had 512 agents whose parameters were sampled as in equations \eqref{eq:mu} and \eqref{eq:cov}. Note that the two figures have different time-scales to illustrate how the parameters of the model (see \S\ref{sec:twostate}) affect the oscillations between staked and lent supply}
    \label{fig:lend_cross}
\end{figure}

\begin{figure}
    \centering
    \includegraphics[scale=0.38]{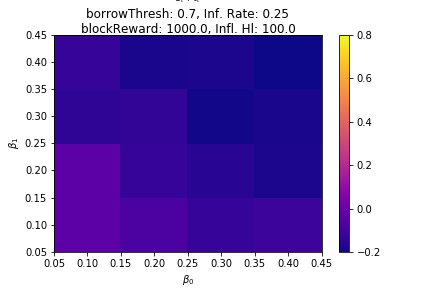}
    \includegraphics[scale=0.38]{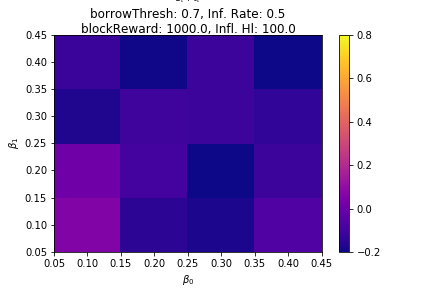}
    \includegraphics[scale=0.38]{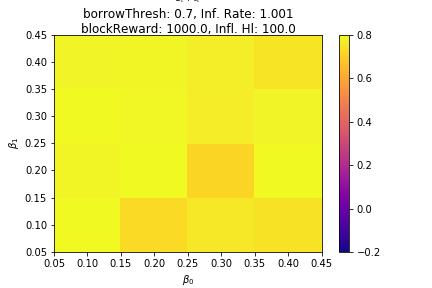}
    \includegraphics[scale=0.38]{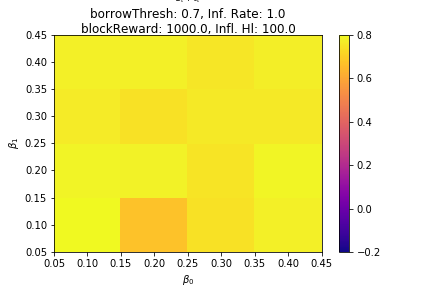}
    \caption{Plots of relative amount of staking to lending, with inflation increasing in a clockwise manner. When these plots are negative, there is, on average, more lending than staking and vice-versa, when positive. When we are deflationary (top two rows), we see that regardless of the lending rate parameters (the x-axis is $\beta_0$ and the y-axis is $\beta_1$ from equation \eqref{eq:bonding_curve}), tokens on average lent. This empirically demonstates that there is a phase transition from a lending-dominated regime (which is deflationary) to one that is staking-dominated and inflationary. This simulation also uses 512 agents and the titles state the inflation rate (which is deflationary if it is less than $1$), block reward, half-life, and borrow threshold (see \S\ref{sec:bdd}) and each sample involves an average over a hundred random seeds.}
    \label{fig:phase_transition}
\end{figure}

\begin{figure}
    \centering
    \includegraphics[scale=0.38]{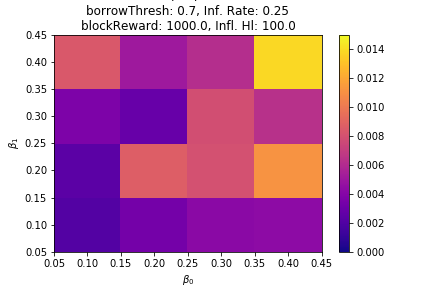}
    \includegraphics[scale=0.38]{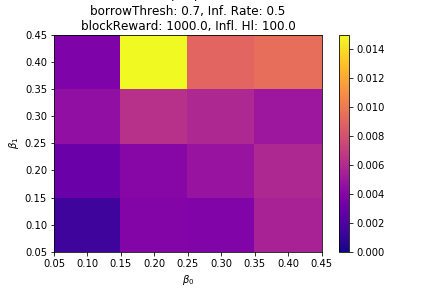}
    \includegraphics[scale=0.38]{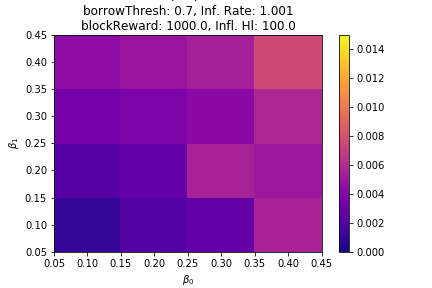}
    \includegraphics[scale=0.38]{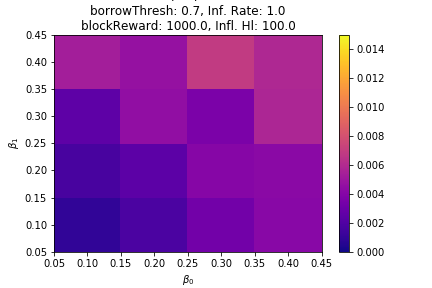}
    \caption{Plots of the relative spread between borrowing and lending rates. If the value is higher, then borrowing costs more than lending. Note that in the deflationary regime, as the lending rates increase (e.g. as $\beta_0, \beta_1$ from \eqref{eq:bonding_curve} increase), the spread also increases, whereas this effect is far more muted. The simulation parameters are the same as figure \ref{fig:phase_transition}.}
    \label{fig:spreads}
\end{figure}

We answer these questions with a stochastic model that can be theoretically solved in certain situations and is easily simulated via Monte Carlo methods.
Our aim is not to model realistic networks parameters perfectly, but rather to show that even in the most simplified model of agents optimizing portfolios composed of staked and lent tokens, on-chain lending can cause dramatic volatility in network security.
In particular, we construct an agent-based model, where each network participant is represented via an agent with a utility and decision function.
Agent-based modeling has been previously used for modeling censorship properties in sharded PoW chains \cite{chitra2019agent} and can serve as a conduit for comparing theoretical results to empirical data in a statistically rigorous manner.
Our model for rational network agents involves having each participant view their total token wealth as a two component portfolio of tokens staked and lent.
We assume that the agents have different risk preferences and locally optimize their token portfolios using mean-variance optimization \cite{markowitz1952portfolio}, which allows for agents to adjust their portfolios based on observed returns and risk preferences. 
Figure \ref{fig:phase_transition} illustrates that there is a phase transition in inflation rate that leads to tokens going from being predominantly lent in deflationary regimes to predominantly staked in inflationary regimes.
Moreover, figure \ref{fig:spreads} shows that the spread between borrowing and lending rates is significantly worse for deflationary PoS assets, further confirming the existence of a phase transition. 
We explicitly state our simplifying assumptions in \S\ref{sec:ass} and note that agent-based modeling allows one to relax these assumptions and analyze how these results carry over to real PoS protocols, such as Cosmos \cite{buchman2019byzantine} and Tezos \cite{goodman2014tezos}.
In section \S\ref{sec:formal}, we prove properties of this model that match the observed phase transition from figures \ref{fig:phase_transition} and \ref{fig:heatmap}.
A single sample path is depicted in Figure \ref{fig:lend_cross}, which visually show that agents participating in on-chain lending and staking can cause a ``flippening," where there are more assets lent out than staked.
The observed volatility in the amount of assets staked is tantamount to dramatically reducing the cost of taking over a staking network, which implies that the security model of PoS networks needs to account for attacks on the network that stem from reduced cost of capital.
The combination of theoretical and simulation-based results demonstrate that the threat model for PoS networks needs to be expanded to include financial attacks that result from yield competition with on-chain financial products.

For provenance, note that the mathematical notation used in this paper is documented in Appendix \ref{app:notation}.

\section{Protocol Assumptions}\label{sec:ass}
In order to simplify our model, we will make some simplifying assumptions. 
These assumptions hold for all models analyzed in this paper and they focus on properties of the underlying distributed ledger rather than properties about the economic behavior of participants, which is described in the proceeding section.
\begin{assumption}[Cryptographic]
All sampling processes use true randomness and not pseudorandomness 
\end{assumption}
This differs from many standard cryptographic threat models that assume pseudorandomness and provide an $\epsilon$-approximation to true random sampling \cite{koblitz2015random}.
In the Appendix, we use this property to ensure that our PoS algorithm is non-anticipating (e.g. adapted to a suitably chosen filtration), allowing for conditional probabilities to be computed without having to aggregate the $\epsilon$ error terms.
One can remove this assumption, in a similar manner to \cite{chan2018pala}, with a significant increase in proof complexity.

\begin{assumption}[Distributed Systems]
All communication between participants is synchronous. 
\end{assumption}
This can be relaxed to partial synchrony, with an increase in complexity in the proofs and simulations described in the following section.
\begin{assumption}[Identity]
All pseudonymous identities are known by all participants and the number of participants (measured by unique addresses), $n$, is fixed.
\end{assumption}

\begin{assumption}[Time]
The entire system will update at discrete time intervals, with each tick to be thought of as a block update. 
\end{assumption}
While participants will likely execute strategies in continuous time, assuming discrete time evolution ensures that participants only respond to event updates that are received on-chain. 
One can also remove this assumption at the cost of increased variance using `Poissonization' techniques \cite{kane2016new} and `sleepy consensus' assumptions \cite{pass2017sleepy}.

\begin{assumption}[No External Markets]
Only on-chain lending, borrowing, and staking will be considered.
\end{assumption}
We are ignoring off-chain lending (e.g. OTC desks or lending businesses, such as Galaxy and Tagomi) and are assuming that there always exists block space for any participant's action to succeed. We are omitting this to reduce the complexity of our model, as off-chain lending has varied pricing models and term structures.

\begin{assumption}[Money Supply is Deterministic]
The block reward at time $t, R_t$ and the money supply at time $t, S_t = \sum_{s=0}^{t} R_t$ are deterministic and known to all participants.
\end{assumption}
In particular, we avoid assuming that there exist governance mechanisms for changing the block reward, which have been proposed for protocols such as Algorand \cite{gilad2017algorand} and Celo \cite{kamvarcelo}. 

\begin{assumption}[No Transaction Fee Revenue]
The only revenue that validators receive from staking comes from the block reward.
\end{assumption}
This is a model assumption that removes the complexity of modeling crypto network fee markets, which currently have unstable dynamics and are poorly understood.

\begin{assumption}[PoS]
The following properties hold for our idealized PoS algorithm:
\begin{itemize}
    \item \emph{No compounding}: PoS mechanism uses epoch-based sampling and does not immediately reinvest block rewards. This is to avoid the concentration behavior described in \cite{fanti2019compounding}. One can relax our model to handle P\'olya urn processes, at the cost of significantly worse variance.
    \item \emph{Single validator per block}: To simplify the model, we avoid using committees \cite{gilad2017algorand,rocket2019scalable,kamvarcelo,goodman2014tezos,buchman2019byzantine,chitra2019committee,pass2018thunderella,neuder2019selfish} and verifiers \cite{bagaria2018deconstructing,baudet2018state} as they add more variance and make both formal and simulation methods more difficult.
    \item \emph{Constant slashing probability}: We assume that each validator has a fixed probability $\slashprob$ of being slashed on a block that they produce. This is simplistic as it assumes that all validators have the same chance of being slashed, regardless of stake and validation history. However, in practice, we have seen very few live slashes and this model encourages simpler formal and simulation analysis.
    \item \emph{Sharded state is synchronously traversed} Any sharded state in our PoS blockchain is read synchronously and we assume linearizability in our blockchain. This can be relaxed to non-linearizable protocols like Avalanche \cite{rocket2019scalable}, but will depend on the scoring function used for each branch.
    \item \emph{No unbonding period} We do not assume that the PoS protocol has an unbonding period like that of Cosmos \cite{buchman2019byzantine} or Tezos \cite{goodman2014tezos}.
    \end{itemize}
\end{assumption}

\section{Model}
\subsection{PoS Model}
Let $S_t$ be the total outstanding token supply of a PoS protocol at time $t$ and let $S_t = \zeta_t + \ell_t$, where $\zeta_t$ is the number of staked tokens and $\ell_t$ is the number of lent tokens. We use a simple model of a PoS system that samples a single block producer from a discrete time-series of stake distributions, denoted by $\stakedist(t) \in \zeta_t \Delta^n$, where $\Delta^n$ is the $n$-dimensional probability simplex (see Appendix \ref{app:notation}). The main input parameters are:
\begin{itemize}
    \item $\stakedist(0)$: Initial asset distribution
    \item $R_t$: Staking block reward at block height $t$
    \item $S(i, \stakedist)$: Slash that validator $i$ receives if the staking distribution is $\stakedist$
\end{itemize}
The formal specification of the algorithm can be found in Algorithm \ref{alg:pos_sim} in Appendix \ref{app:pos}. The algorithm state includes the current stake distribution, the current epoch's reward set, the current epoch's slash set, and the current block time. At a high-level, for each block, we select a validator who should receive a block reward and decide if they are to be slashed by flipping a coin with probability $\slashprob$. If they are slashed, we add their \emph{id} and the amount that they are to be slashed to the current epoch's slash set. Otherwise, we add them to the block reward set. The algorithm updates the stake distribution on a per epoch basis.  

\subsection{Lending Model}
We will study two models for lending, one involving an explicit model for borrowing demand and one that is implicit.
The implicit model, which is simpler, provides a prototype for constant borrowing demand and is amenable to theoretical results.
On the other hand, the model involving an explicit description of borrowing demand is more realistic and amenable to being fit by historical data.

\subsubsection{Two-State Lending Model}\label{sec:twostate}
In the two-state lending model, we track the time evolution of two token distributions, $\stakedist(t), \lenddist(t) \in S_t \Delta^n$, which respectively represent the distribution of staked tokens and those locked in a lending contract.
The $i$th component of the stake distribution, $\stakedist(t)_i$, corresponds to the amount of tokens that the $i$th agent has staked.
Each agent has a wealth $W_i(t)$ at time $t$ that is equal to the sum of their portfolio of staked and lent tokens, i.e. $W_i(t) = \stakedist(t)_i + \lenddist(t)_i$.
By definition, the total lending supply is equal to the sum of all lent portfolios, i.e. $\ell_t = \Vert \lenddist(t) \Vert_1$, and the total money supply is equal to the sum of all portfolios, i.e. $S_t = \Vert \stakedist(t)\Vert_1 + \Vert \lenddist(t) \Vert_1$.
At each time step, agents update their portfolios based on returns accrued from the previous time step and after portfolios are updated, the lending rate, $\gamma_t$, is updated based on the total amount lent. This means that we are making two implicit assumptions:
\begin{enumerate}
    \item \emph{Constant relative borrowing demand}: We are assuming that ratio of borrowing demand (represented via a quantity of tokens) to the total token supply stays constant, as the rate only depends on the lending and staking supplies. Formally, this means that the demand at time $t$ is equal to $kS_t$
    \item \emph{Flows are the only determining factor}: Participants who move tokens from staking to lending or vice versa are the only causes for changes to the lending rate
\end{enumerate}
We draw inspiration from Compound \cite{compound2019}, which provides a simple formula for the borrow lending rate, $\beta_t$, and the lending rate, $\gamma_t$. 
The Compound model computes a utilization rate $U_t$ at block height $t$, which is the ratio of the borrowing demand to the token supply locked in the contract, and uses that to update formulas for $\beta_t, \gamma_t$. Mathematically, they define the utilization rate as
\[
U_t = \frac{k S_t}{\ell_t + k S_t}
\]
We compute the borrow and lend rates using the following formulas, where $\beta_0, \beta_1 \in (0,1)$ are interest-rate parameters and $\gamma_0 \in (0,1)$ is a measure of the spread between lending and borrowing (i.e. $1-\gamma_0$ is the relative spread).
\begin{align}\label{eq:bonding_curve}
    \beta_t &= U_t(\beta_0 + \beta_1 U_t) \\
    \gamma_t &= (1-\gamma_0)\beta_t 
\end{align}
For reference, the Compound V2 contract uses the values $\beta_0 = 5\%$ and $\beta_1 = 45\%$. 
As depicted in Figure \ref{fig:lend_cross}, there can be an enormous amount of volatility in the fraction of the token supply that is lent, $\frac{\ell_t}{S_t}$. 
In \S\ref{sec:formal}, we prove tail bounds on the inflows and outflows of lent tokens over a time step, e.g. $\Prob[|\ell_t - \ell_{t-1}|> \epsilon S_t]$, that explicitly depend on the block reward, $R_t$ and the interest rate parameters.
These bounds suggest that even for the overly simplistic setting of the two-state model, PoS protocols need to carefully choose their block rewards if they desire to have a large fraction of the outstanding token supply staked at all times.\footnote{It should be noted that this has been a stated goal of large pools of validators and the existence of staking services that pool tokens and operate nodes exemplifies this collective desire to have a large portion of the token supply staked \cite{figment_tezos,iqlusion}}

\subsubsection{Three-State Model}
Instead of assuming that there is constant relative borrowing demand, we can relax this by specifying an additional distribution, $\borrowdist(t)$, such that $\tokendist = \stakedist + \lenddist + \borrowdist$ and $S_t = \zeta_t + \ell_t + \xi_t$, where $\xi_t = \Vert \borrowdist(t)\Vert_1$ is the total amount borrowed at time $t$. In this world, the utilization ratio is now defined as:
\[
U_t = \frac{\xi_t}{\ell_t + \xi_t}
\]
Formally analyzing this model has a variety of difficulties that stem from that fact that we have to explicitly model the borrow demand distribution and disentangle how it couples to each participant's local model of risk, which is defined in the next section. In \S\ref{sec:abs}, we simulate this model with a variety of different borrowing demand distributions, but all formal proofs that follow only analyze the two-state lending model. Since the pool of borrowers (e.g. arbitrageurs) is often disjoint from the pool of stakers and lenders, we will make the following assumption: 
\begin{assumption}[Borrowing demand is independent of staking and lending]\label{as:bd}
The borrowing demand distribution $\borrowdist$ is probabilistically independent of the lending and staking distributions.
\end{assumption}

\subsection{State Transition}\label{sec:statetrans}
The final task needed to completely specify this model is to define the state transition rule that sends $\lenddist(t), \borrowdist(t), $ $\tokendist(t)$ to $\lenddist(t+1), \borrowdist(t+1), \tokendist(t+1)$.
As per Assumption \ref{as:bd}, the evolution of $\borrowdist(t)$ is independent of staking and lending, and will be specified separately in \S\ref{sec:abs}.
We therefore need to specify state transition rules on a per agent basis, where each agent's state is their token portfolio $(\stakedist(t)_i, \lenddist(t)_i)$.
Traditionally, the strategy space of rational actors is described via an expected utility function that an agent aims to maximize by taking various allowable actions.
Before specifying the strategy space that we will sample, let's consider a few examples to motivate the need for agents who have varying risk preferences.
If at time $t$, staking is returning more than lending and every agent moves their entire portfolio from lending to staking, then we will observe a correlated spike in lending rates $\ell_t$ will go to zero and $U_t = 1$.
Moreover, the relative staking reward to any agent will decrease as $S_t$ will increase by the amount of tokens that flow from lending to staking as the expected return on an epoch for an agent is their staked tokens divided by the token supply, $\frac{\stakedist(t)_i}{S_t - \ell_t}$.
Thus the greedy strategy of moving all of one's assets to the higher yielding activity and causes drastic swings in the relative yields of staking and lending is unstable and doesn't accurately represent reality, where token holders have differing risk preferences and will not immediately move their entire portfolio from staking to lending (or vice-versa).
Furthermore, cryptocurrency holders are often looking for returns that are multiples of their initial investment and have a long time-preference \cite{weber2016bitcoin}.
In our evolution of each agent's portfolio, we also assume that an agent making a decision at time $t$ can only use the information about all portfolios up to time $t$, $\{\pi_i(t) : i \in [n]\}$, and the implied rate $\gamma_t$.
Since we are dealing with on-chain lending only, this assumption says that players cannot use strategies that look into the future and that all agent portfolios are public.
In order to capture strategies that are independent of front-running and latency arbitrage, we will make the following assumption:
\begin{assumption}[Martingale Ordering]\label{as:martingale}
There exists an $[n]$-valued martingale $Z_t$ that chooses the ordering in which participants are allowed to update their portfolios at time $t$ to those at time $t+1$.
\end{assumption}
Under this assumption, agents receive no advantage in expected returns by trying to predict when their strategy is executed (e.g. is agent 1's strategy executed before agent 2's strategy because agent 1 has more staked than agent 2?).
This assumption is reasonable as our goal is to figure out if rational, but non-Byzantine, agents will cause PoS network security to decrease when on-chain lending activity is sizeable.

\subsubsection{Optimal Portfolio Construction}
Mean-variance methods, pioneered by Markowitz's Nobel Prize winning work on portfolio theory \cite{markowitz1952portfolio}, provide a way for rational traders of risky assets to construct portfolios that trade-off individual preferences for maximizing returns with those of risk minimization.
These methods, which are the backbone of the majority of trillions of dollars of passive portfolios and statistical arbitrage strategies, provide a simple, easy-to-solve model that involves two parameters for constructing portfolios of $n$ assets: an expected return vector $\mu \in \mathbb{R}^n$ and a positive-definite covariance matrix $\Sigma \in S_+^{n\times n}$.
Given these parameters, one solves a strongly convex program that aims to compute the fraction of an agent's wealth that should be allocated to each asset while ensuring that the sum of the allocations is one and each entry is positive.
In particular, the seminal work of Markowitz aimed to optimize the quadratic form $f : \R^n \times \R^n \times \R \times S_+^{n\times n} \rightarrow \mathbb{R},\; f(w, \mu, \lambda, \Sigma) = w^T \Sigma w - \lambda \mu$, where $\lambda$ is a parameter that controls the riskiness of the output portfolio and $w$ is the portfolio allocation.
As $\lambda$ is varied, the \emph{efficient frontier} of admissible portfolios is defined as the surface $S(\mu, \Sigma) = \{w \in \R^n : \exists \lambda \text{ such that } w \in \argmin\, f(w, \mu, \Sigma, \lambda)\}$.
The original work of Markowitz \cite{markowitz1952portfolio} focused on the single period allocation problem, where an investor aims to find the optimal portfolio over a single time-period, which corresponds to assuming that $\mu$ and $\Sigma$ do not change over time.
Further work on multiple period \cite{li2000optimal} and continuous-time methods \cite{zhou2003markowitz} for mean-variance optimization allow for $\mu$ and $\Sigma$ to vary as functions of time, with the continuous-time methodology drawing $\mu$ from an \^Ito process, such as a solution to the Black-Scholes equation for options pricing.
As blockchain systems have incremental updates with independent games per update (e.g. transaction fee markets can differ wildly from block to block), we will necessarily have to consider the multiple period model and define how each agent's mean vector and risk-preferences evolve over time. 
Finally, we note that our methodology is directly comparable to that of multistrategy backtesting in quantitative trading \cite{boyd2017multi}.

We will assume that each agent treats their token wealth, $(\lenddist(t)_i, \stakedist(t)_i)$ as a Markowitz portfolio and updates, on receiving tokens from staking and lending, an estimate for a time-dependent return vector $\mu_i(t) \in \mathbb{R}^2$.
We also assume that each agent has a different, time-independent covariance matrix $\Sigma_i$ that is drawn from a random matrix ensemble\footnote{In graphical models and probability, such as the Ising model or spin glasses, this is known as ``quenched disorder" \cite{mezard2009information}}.
In other words, the expected return vector adjusts with time (it depends on the staked and lent quantities), while the covariance stays fixed in time.
As long as there is some variance in the chosen random matrix ensemble (e.g. $\exists i, j \in [n]$ such that $\mathsf{Pr}[\Sigma_i \neq \Sigma_j] > 0$), then the dynamics will not deadlock into a state where all participants end up with the same portfolio (e.g. all outstanding tokens are 100\% allocated to staking or lending).
If agents simply move all of their assets from one pool to another, as opposed to some risk-adjusted proportion, then the system can deadlock quickly when borrowing demand is constant.
Using the notation of the prequel, we  define $\mu_i(t) = \mu_i(\stakedist(t), \lenddist(t))$ as:
\begin{equation}\label{eq:mu}
    \mu_i(t) = 
        \left[\begin{array}{c}
             \mustake(t)_i
             \\
         \mulend(t)_i
    \end{array}\right]
    =
    \left[\begin{array}{c}
          \frac{\stakedist(t)_i}{S_t - \ell_t}   \\
         \gamma_t 
    \end{array}\right]
     = 
         \left[\begin{array}{c}
          \frac{\stakedist(t)_i}{\zeta_t}   \\
         \gamma_t 
    \end{array}\right]
\end{equation}
Let us motivate this choice of expected return vector. 
Recall that the return vector is supposed to represent the relative rate of return (usually referred to as an alpha in the quantitative finance literature) over a riskless asset. 
In this situation, the riskless asset is holding our tokens (as bearer instruments) and we expect to earn 
yields\footnote{For staking, the relative yield is defined as future expected staked wealth divided by current stake wealth minus 1, which is
\[
\frac{\stakedist(t)_i + R_t \frac{\stakedist(t)_i}{S_t-\ell_t}}{\stakedist(t)_i} - 1 =  \frac{R_t}{S_t - \ell_t}
\]
This is the standard form of how yields are specified in traditional Markowitz portfolio theory \cite{markowitz1952portfolio}.
}
of $\frac{R_t}{S_t-\ell_t}$ and $\gamma_t$ for staking and lending, respectively.
However, note that the former yield can be greater than 1 (when $S_t - \ell_t < R_t$), whereas the latter yield cannot as per equation \eqref{eq:bonding_curve}.
In order for Markowitz optimization to be well-defined, we need to choose yields that are directly comparable (e.g. have the same range).
Moreover, we note that the na\"ively calculated staking yield loses the dependence on the $i$th party's current wealth.
Since our system allows for slashing and each validator incurs a variance in reward proportional to the inverse square root of the epoch length, it is preferable to find a yield that depends on the $i$th parties wealth to reflect this variance.
The simplest estimator for a staking validator's yield that is in $[0,1]$ (comparable to lending yield) and has a component of the variance is the probability of winning the block reward, $\frac{\stakedist(t)_i}{S_t - \ell_t}$, which is exactly what \eqref{eq:mu} describes.

In order to describe the covariance matrix, we will first state an assumption that aims to connect the variances in the system with the expected staked and lent times:
\begin{assumption}\label{as:static_cov}
 The $i$th agent's covariance matrix $\Sigma_i$ will be static (e.g. does not vary with block height), diagonal, and will have variances connected to the expected stake time $\staketime$ and expected lent time $\lendtime$
\end{assumption}
Formally we use the following model for the covariance matrix:
\begin{align}\label{eq:cov}
    \Sigma^{-1}_i &= \left[\begin{array}{cc}
        \alpha_i & 0  \\
         0 & \beta_i
    \end{array}\right]  \\
    \alpha_i &\sim \mathsf{Exp}\left(\staketime\right) \\
    \beta_i &\sim \mathsf{Exp}\left(\lendtime\right)
\end{align}
One can interpret $\staketime$ as the expected epoch length (see Algorithm \ref{alg:pos_sim}), while $\lendtime$ represents the withdrawal window under which a lender can remove their tokens from the lent pool.\footnote{This is implemented in a manner analogous to how hedge funds only let you recall your invested capital in small chunks over a window of time}
By connecting the covariance matrix to these quantities, we encode the connection between a validator's expected risk preference and the time that capital is locked into either staking or lending.
Equation \eqref{eq:cov} represents the fact that risk and time preferences are, to first-order, inversely correlated (e.g. you are willing to lock up capital for the longest duration in the least risky assets) \cite{de1997international}. 
To explain these choices of random variables, first recall that the Markowitz objective function is of the form $f(\mu, \Sigma, w, \lambda) = w^{T} \Sigma w - \lambda \mu$, where $\lambda$ parametrizes the agent's preference for return maximization over risk minimization.
If we divide this objective function by $\lambda$, provided that $\lambda > 0$, we get an equivalent optimization problem for $\lambda^{-1}f$.
Note that $\lambda$ can be thought of as encoding 'collective' risk-preferences for all agents (e.g. a `bull market' when $\lambda$ is large, a `bear market' when $\lambda$ is small).
As such we need a methodology for choosing $\lambda$ that depends on the network size and/or number of agents.
We make $\lambda \sim \chi^2(n)$ in order to capture the fact that as the number of participants increases, so should the expected number that are risk seeking.
Once we do this, we can directly interpret the variables $\alpha_i, \beta_i$ as encoding both this risk preference and the inherent duration difference between lending risk and staking risk.
The static nature of the covariance matrix in Assumption \ref{as:static_cov} implies that each agent has an `equilibrium' risk-preference.
Moreover, this implies that the distribution of risk-preferences amongst changes is not changing --- which one can expect in the limit as the number of agents goes to 
infinity.\footnote{The easiest way to see this is to note that one expects a central limit theorem (e.g. Marchenko-Pastur law \cite{tao2012topics}) for the distribution of $\frac{1}{\sqrt{n}}\sum_{i=1}^n \Sigma_i$. Thus, as the number of participants goes to infinity, we get closer and closer to having the staticity assumption hold true}  
Making this assumption allows for significantly faster simulation (e.g. less compute needed to reduce the variance of the estimator below an error threshold $\epsilon$) 

To summarize, we evolve the system by having each agent update their Markowitz estimate at each time step, which changes the lending and staking distributions for the next time step.
In simulation, agents will only be able to migrate their staking tokens at times $k\staketime, k \in \N$. Explicitly, we evolve the system via the following loop:
\begin{itemize}
    \item Initialize distributions $\{\stakedist(0)_i\}_{i\in[n]}, \{\lenddist(0)_i\}_{i \in [n]}$.
    \item Initialize empirical distributions $\{\estakedist(0)_i\}_{i\in[n]}, \{\elenddist(0)_i\}_{i \in [n]}$
    \item For $t = 1, 2, 3, \ldots$ 
    \begin{itemize}
        \item Observe empirical distributions $\estakedist(t)_i, \elenddist(t)_i$ \\ (generated via actual staking / lending rewards accrued in the epoch)
        \item Compute Markowitz weights $w_{i,t} = (p, 1-p), p = p(\estakedist(t)_i, \elenddist(t)_i, \alpha_i, \beta_i) \in [0, 1]$
        \item Define new portfolio as $p_i(t) = p(p_i(t-1) + \ell_i(t-1))$, $\ell_i(t) = (1-p)(p_i(t-1) + \ell_i(t-1))$ 
    \end{itemize}
\end{itemize}

\section{Formal Properties}\label{sec:formal}
We will describe a few formal properties (proved in the appendix) of the two-state model, as it is feasible to analytically analyze this model.
Our goal in this section is to bound the amount of turnover in the stake distribution that is caused by on-chain lending becoming more attractive than staking rewards to rational stakers.
A secondary goal is to understand what properties of the stochastic processes that represent the evolution of the lending and staking distributions are necessary and/or sufficient for ensuring that we do not have a lot of volatility in agent token portfolios.
This is important as volatility in these portfolios implies that PoS networks have volatile security, which is a distinct defect when compared to PoW.
We will make a few additional assumptions that are necessary to provide analytical results:

\begin{assumption}[Bounded Size]\label{as:bd_sz}
There is a minimum fraction $\delta > 0$ of the money supply that needs to be staked, e.g. $S_t - \ell_t > \delta S_t$ 
\end{assumption} 
If no one is staking then the on-chain lending contract has no value (as it doesn't have any security), so this is a realistic assumption that matches the practical parameters chosen in live networks such as Tezos \cite{goodman2014tezos} and Cosmos \cite{buchman2019byzantine}.
Note that one can directly interpret $\delta$ as the fraction of altruistic validators who will never reallocate or rebalance their assets.

\begin{assumption}[Number of Agents]\label{as:num_agents}
 The number of agents $n$ is larger than a constant multiple of the product of the exponential parameters, e.g. $n = \Omega(\staketime \lendtime)$. 
\end{assumption}
This assumption is required for purely technical reasons that are explained in the proofs in the appendix.\footnote{Also note that the equilibrium assumption (Assumption \ref{as:static_cov}) implicitly also requires a lower bound on $n$ that is a function of $\staketime$ and $\lendtime$ in order for the covariance matrix distribution to be $\epsilon$ away from the limit distribution} 
If we solve the \emph{unconstrained} problem\footnote{In the constrained version of the problem needs to satisfy Karush-Kuhn-Tucker conditions \cite{mangasarian1994nonlinear} that enforce $w^{T} 1 = 1$ and $w_i > 0,\,\forall i$. For simplicity of exposition, we elided these during this section (even though these conditions are satisfied during simulation).} for the Markowitz objective function, $f(\mu, \Sigma, w, \lambda) = w^{T} \Sigma w - \lambda \mu$, then $w$ needs to satisfy the first-order condition, $\nabla_w f(\mu, \Sigma, w, \lambda) = 0$.
This yields,
\begin{equation}\label{eq:markowitz_update_rule}
    w = \lambda\Sigma^{-1} \mu
\end{equation}
Given that we are in the multi-period Markowitz setting, this means that we can estimate the $i$th participant's portfolio weights via $w_i(t) = \lambda \Sigma_i^{-1}\mu_i(t)$.
As a measure of volatility in the security of the underlying staking mechanism, we will first look at how $w_i(t)$ changes in time.
Intuitively, this change corresponds to how large rebalancing events (e.g. moving tokens from staking to lending, or vice-versa) are between subsequent blocks.
If this rebalancing is large, then the network could dramatically reduce it's security as holders move their assets from staking to lending.
On the other hand, if this rebalancing is small and decreases over time, then we know that the staked token supply is stable. 
Using the Markowitz update rule \eqref{eq:markowitz_update_rule}, we have the following bound.
\begin{equation}
    \Vert w_i(t+1) - w_i(t) \Vert_1 = \Vert \Sigma^{-1} (\mu_i(t+1) - \mu_i(t)) \Vert_1
    \leq \Vert \Sigma^{-1}\Vert_{1\rightarrow 1} \Vert \mu_i(t+1) - \mu_i(t) \Vert_1 
\end{equation}
where $\Vert A \Vert_{1\rightarrow 1} = \max_{v\in\mathbb{R}^n, \Vert v\Vert_1 = 1} \Vert Av\Vert_1$ is the $L^1$ operator norm.
This simple inequality implies that the volatility in portfolio weights, represented by the single block difference in weights, is controlled by the difference in the mean vector, as $\Expect[\Vert \Sigma^{-1}\Vert_{1\rightarrow 1}] = \Expect[\alpha \vee \beta] \leq \staketime \vee \lendtime$.
Therefore, we focus on trying to bound the difference in expected returns as a function of the block reward at time $t$, $R_t$, the total token supply $S_t$, the lent supply $\ell_t$, and the Compound lending curve parameters $\beta_0, \beta_1$.

Let $\stakedist(t+1)_i - \stakedist(t)_i = \dstake(t)$ and $\ell_{t+1} - \ell_t = \dlend(t)$.
Since the PoS algorithm is adapted to a filtration $\mathcal{F}_t$ on $S_t \Delta^n$ and the covariance matrices are constant as time evolves,\footnote{In particular, they are random at time $t=0$ and defined in $\mathcal{F}_0$, but are constant throughout. This is a measure theoretic way of defining quenched disorder \cite{talagrand2010mean}} $\stakedist(t), \lenddist(t)$ are also $\mathcal{F}_t$-adapted random variables. 
Note that by definition, $\dstake(t) \leq S_t$ and $\dlend(t) \leq S_t$, since we cannot change the amount staked or lent by more than the outstanding money supply. We will first bound the difference in expected returns as a function of $\dstake(t), \dlend(t), $ and $S_t$:
\begin{claim}\label{claim:stake_diff}
There exist constants $C, C' > 0$ such that $$\Vert \mu_i(t+1) - \mu_i(t)\Vert_1 < \frac{C|\dstake(t)|}{S_{t+1}} + C'\dlend^2(t)$$
\end{claim}
Note that these constants are allowed to depend on $\delta$ from Assumption \ref{as:bd_sz}.
If we have an compounding and inflationary rewards schedule, e.g. $S_t = e^{\lambda t}$, then this claim implies the following:
\begin{claim}\label{claim:stake_bd}
If there exists $\lambda > \staketime$ such that $S_t = \Omega(e^{\lambda t})$ and $\min_i W_i(t) > 0$, then as $t \rightarrow \infty$, the maximum change in stake is bounded above by the lending volatility, $\dlend^2(t)$, w.h.p.
\end{claim}
This claim implies that if there is not much variance in the lending rate, either due to choosing small parameters $\beta_0, \beta_1$ or because borrowing demand is minimal, then we should not expect portfolios to rebalance regularly and rational stakers will tend to keep their tokens locked in a staking contract.
Another natural quantity to look at is the variance of the lent assets.
We show that the money supply and time-preference for lending, $\lendtime$, control the variance of lent assets.
\begin{claim}\label{cl:var}
Let $\mathcal{F}_t$ be the filtration such that the lending process $\ell_t$ is adapted. Then we have:
\[
    \Var[\ell_{t+1} \vert \mathcal{F}_t] = \frac{\gamma_t}{\lendtime^2} \Vert W(t) \Vert_2^2
\]
Moreover, we have the following bounds:
\[
\frac{\gamma_t^2 S_t^2}{\lendtime^2 \sqrt{n}} \leq  \Var[\ell_{t+1} \vert \mathcal{F}_t] \leq \frac{\gamma_t^2 S_t^2}{\lendtime^2}
\]
\end{claim}
Note that if $k$ is the constant representing the ratio of borrowing demand to $S_t$ (c.f. \S\ref{sec:twostate}), there exists a constant\footnote{Since $U_t = \frac{kSt}{\ell_t + kS_t}$ and Assumption 12 implies that $\ell_t < (1-\delta)S_t$, we have $U_t > \frac{k}{k + 1 - \delta}$ and $\gamma_t > \beta_1 \left(\frac{k}{k+1}\right)^2 + \beta_0\frac{k}{k+1}$.} $\aleph$ that depends on $k$ such that $\gamma_t \geq \aleph$.
As such, Claim \ref{cl:var} implies that as long as $S_t = \Omega\left(n^{\frac{1}{4}}\right)$, we have reallocation from staking to lending.
Thus, any monetary policy that grows sufficiently quickly with the number of users of the network will \emph{always} have assets moving into and out of lending.
If we place constraints on the demand $k$, we can strengthen this result into a bound on how much $\dlend^2$ oscillates:
\begin{claim}\label{cl:submart}
Let $\eta_t = 1 + \frac{\Vert W(t)\Vert_2^2}{S_t^2} \geq 1 + \frac{1}{\sqrt{n}}$ and $\alpha_t = \frac{\ell_t \lendtime}{S_t \eta_t}$ If for all $t$, $k \geq \frac{\alpha_t}{1 - \alpha_t}$, where  and the hypotheses of claim \ref{claim:stake_bd} hold, then $\dlend^2(t)$ is a submartingale and we have
\[
\Prob\left[\max_t \dlend^2(t) > \lambda\right] < \frac{1}{\lambda^2}\mathsf{Var}[\dlend^2(t)]
\]
and subsequently,
\begin{equation}\label{eq:tailb}
    \Prob\left[\max_t\Vert \mu_i(t+1) - \mu_i(t)\Vert_1 > \lambda\right] < \frac{1}{\lambda^2}\mathsf{Var}[\dlend^2(t)]
\end{equation}
\end{claim}
In words, this claim says that as long as we have inflation and there is enough borrowing demand, then we can be sure that the \emph{worst-case} rebalancing is bounded by the variance of lending volatility.
If we add another constraint on the behavior of the increments $\dlend^2(t)$, then we can strengthen this claim to get a phase transition that resembles the Galton-Watson phase transition \cite{asmussen1983branching}.
\begin{claim}
Suppose that for all $t > 0, \dlend(t) < \frac{\ell_t^2}{2 \ell_{t-1} \eta_t}$, $\Expect[\dlend(t)] > 0$, and let $\eta_t$ be as in Claim \ref{cl:submart}. Define $r_{\pm}$ as follows:
\[
r_{\pm} = \frac{\ell_t \lendtime}{S_t \eta_t} \left(1 \pm \sqrt{1 + \frac{\eta_t \ell_{t-1}(\ell_{t-1}-2\ell_t)}{\ell_t^2}}\right)
\]
Then $\ell_t^2$ is a supermartingale when $\gamma_t \in (r_-, r_+)$, a submartingale when $\gamma_t \in [0, r_-) \cup (r_+, 1]$, and a martingale when $\gamma_t \in \{r_{-}, r_{+}\}$.
\end{claim}
The intuition for this is as follows:
\begin{itemize}
    \item When there is either too little or too much borrowing demand (e.g. $\gamma_t < r_-$ or $\gamma_t > r_+$), then the expected lent supply either increases (on average) to one or decreases to zero. This is analogous to a gambler's wealth after playing a game with probability $p < 1/2$ for $n$ rounds --- the wealth concentrates into either the house (staked supply) or the gambler (lent supply). 
    \item When there is a moderate amount of borrowing demand, $\gamma_t \in [r_-, r_+]$, then we have stable, potentially oscillatory behavior. Doob's Supermartingale Convergence Theorem \cite{billingsley2008probability} intimates that the distribution is stationary as $t\rightarrow \infty$. This corresponds to a gambler playing a game in which their chance of winning is $p > 1/2$.
\end{itemize}
These results for the simpler two-level model suggest that our simulated phase transition results convey the existence of a deeper phase transition.

\paragraph{Agent-level Behavior} In order to get a stronger understanding of what is going on at the agent level and for different monetary policies, one needs an understanding of the probability that a single agent has a large rebalancing event that affects the staked portion of their portfolio.
This involves studying how the staking components of $\mu_i(t)$ changes over time. 
Let $\mustake(t)_i$ be the different in the staking component of $\mu_i(t)$ between times $t$ and $t+1$.
We can bound the probability that an agent rebalances their portfolio by an $\epsilon$ fraction via the following claim:
\begin{claim}\label{claim:stake}
Let $\mustake(t)_i = \frac{\stakedist(t+1)_i}{S_{t+1} - \ell_{t+1}} - \frac{\stakedist(t)_i}{S_t - \ell_t}$. Then for all $t > 0$, we have
\begin{equation}\label{eq:claim2}
\Prob\left[|\mustake(t)_i| < \epsilon\right | \mathcal{F}_{t-1}] = \Omega\left(1 - \left( \left(\frac{\epsilon S_{t+1}}{\gamma_t}\right)^n e^{-\staketime\lendtime\frac{\epsilon S_{t+1}}{\gamma_t}}\right)\right) 
\end{equation}
\end{claim}
Bounds like \eqref{eq:claim2} are of the form $\Prob[X_t < \epsilon] = Y_t$, which imply that the extrema of $Y_t$ provide a guaranteed bound of how large $X_t$ can be when $Y_t$ is minimized.
This means that we can try to bound the first hitting time $t^*$ for the maxima of $|\mustake(t)_i|$ by analyzing the minima of the right-hand side of \eqref{eq:claim2}.
Note that the function $1-(kx)^n e^{-k'kx}$ has a minima at $x = \frac{n k^{n-2}}{k'}$, which means that we can estimate when, as a function of $S_t, \gamma_t, \epsilon$, we have maximal deviations in stake.
In the following section, we examine this claim for different monetary policies $S_t$.

\subsection{Deflationary Monetary Policy}\label{sec:deflation}
In this case, $R_t = k r^{-t}, r < 1$ and $S_t = \frac{k(1-r^{-t-1})}{1-r} = C_{\infty} - C'r^{-t-1}$, where $C_{\infty}$ is the final money supply (e.g. 21 million, for Bitcoin).
Letting $\epsilon = \delta S_{t+1}$ and plugging this into the right hand side of \eqref{eq:claim2} and optimizing for $t$ using the chain rule gives the condition $t^* = -\log\left(C- \sqrt{\frac{n\delta}{\gamma_t \staketime \lendtime}}\right)$. Requiring that $t^* > 0$ gives the condition $C - \sqrt{n\delta}{\gamma_t \staketime\lendtime} \in (0,1)$ or
\[
\delta > \frac{(C_{\infty} -1)\gamma_t \staketime\lendtime}{n}
\]
The above condition implies that at the time with the highest expected amount of removed stake, we expect that quantity to at be on the order of $\frac{C_{\infty}}{n}$.
If there are fewer terminal coins than participants, $C_{\infty} \ll n$, then we should expect large rebalances for every staker. 
This formalizes the intuition that totally deflationary currencies are vulnerable to large rebalances that depend strongly on the number of coins in the system.
It also confirms the intuition that if there are fewer coins than participants, then borrowing demand will be high and we should again expect large rebalances.

\subsection{Inflationary Monetary Policy}
Next, we consider what Claim \ref{claim:stake} implies about inflationary monetary policies. 
We will consider two types of inflationary policies: Polynomial ($S_t = \Omega(t^k)$) and Exponential ($S_t = \Omega(e^{\lambda t})$).

\subsubsection{Polynomial Inflation}
For polynomial inflation, we have $R_t = c t^{k-1}$ and $S_t = \Omega(t^k)$. Optimizing \eqref{eq:claim2} with this form of $S_t$ gives $t^* = \left(\frac{n\delta}{\gamma_t \staketime \lendtime}\right)^{1/k}$. Plugging this into the right-hand side of \eqref{eq:claim2} gives a lower bound of
\[
1 - \frac{\delta^2 n}{\gamma_t \staketime \lendtime}e^{-\staketime \lendtime\delta^2 n / \gamma_t}
\]
which means that if $\delta = O(n^{-1/2})$, individual agents will not have rebalances of size $\delta S_t$ with high probability.
These guarantees do not depend on the polynomial degree $k$, which means that even simple linear policies ($k=1$) are sufficient to get low stake turnover.

\subsubsection{Exponential Inflation}
In the case of exponential inflation, $S_t = C_0 e^{\lambda t}$ for some $\lambda > 0$ and initial token distribution $C_0$. Following the logic of the \S\ref{sec:deflation} section, we arrive at a conclusion of,
\[
\delta > \frac{C_0\gamma_t \staketime \lendtime}{n}
\]
This states that the worst-case rebalancing is bounded by the \emph{initial token distribution} as opposed to the final token distribution.
Given that most networks assume that there will be more users than the \emph{initial} distribution, $C_0$, this implies that large rebalancing events should become rarer once the network achieves a scale of $n \gg C_0$.

\subsection{Conclusions}
The results of this section show that the model of \S\ref{sec:statetrans} has a volatility that is mainly dependent on lending volatility.
This volatility, if demand is sufficiently shallow, is small enough to help bound the worst-case portfolio rebalancing for any validator in the system.
However, the turnover in staked quantity, measured by how each agent's expected reward changes over each epoch period, is sensitive to the precise nature of the monetary policy, $S_t$.
In particular, deflationary policies cannot support on-chain lending, as the worst-case rebalancing rate depends on the terminal money supply, whereas the rebalancing rate for exponentially inflationary monetary policies only depends on the initial money supply (e.g. Ethereum's pre-mine).
Finally, we note that polynomial inflation provides particularly good rebalancing guarantees that are independent of the rate of growth of the money supply.
Many of these results rely on asymptotic behavior of agents in this system and all of these results depend on simple models for borrowing demand.
The remainder of the paper will focus on using simulation to provide a more realistic picture of \S\ref{sec:statetrans}.

\section{Agent-Based Simulations}\label{sec:abs}
In order to test the three-state model in a quantitatively rigorous and realistic fashion, we turn to agent-based simulations.
One of the main reasons to use agent-based simulation over formal methods is that it hard to formally prove what block reward growth rate is ideal for mitigating volatility in portfolio allocation.
For instance, it is difficult to evaluate whether $R_t = \Omega(t^2)$ or $R_t = \Omega(t)$ provide ideal mitigation for on-chain lending with parameters $\beta_0, \beta_1$.
Moreover, we can test various realistic demand distributions, including ones that are atomic and do not have a probability density.
Prior work on agent-based simulations of blockchain systems \cite{chitra2019agent} has focused on analyzing consensus protocols via event-based simulation.
We follow a similar event-based framework for simulating staking and lending, albeit ignoring the details of peer-to-peer networking and consensus.
Our goal is to sample as many trajectories $X_t \vert \mathcal{F}_t = (\stakedist(t), \lenddist(t), \gamma_t)$ as possible for different combinations of parameters $\beta_0, \beta_1, $ and block reward schedules $R_t$. 

\subsection{Initialization}
In order to generate sample paths $X_t$, we need to specify the following variables:
\begin{itemize}
    \item Initial distributions: $\stakedist(0), \lenddist(0)$
    \item Initial interest rate: $\gamma_0 \in (0,1)$
    \item Interest rate parameters: $\beta_0, \beta_1 \in (0, 1)$
    \item Demand-generating distribution parameters (vary depending on the demand generating distributions chosen): $\mathcal{P}_{\xi}$
    \item Staking and lending time scale: $\staketime, \lendtime$
    \item Slashing probability: $\slashprob$
    \item PRNG seed: $s \in \Z_{2^{64}}$
\end{itemize}

We chose to model $\stakedist(0)_i \sim \mathsf{Exp}(\lambda_{\text{stake}})$ and $\lenddist(0)_i \sim \mathsf{Exp}(\lambda_{\text{lend}})$, as these distributions exemplify the extreme concentration of wealth that accompanies most token distributions.
Exponential distributions are also useful in that the order statistics are also exponential\footnote{If $\stakedist(0)_{(1)} < \ldots < \stakedist(0)_{(n)}$ are the order statistics for the stake distribution, then $\stakedist(0)_{(i+1)} - \stakedist(0)_{(i)} \sim \mathsf{Exp}\left(\frac{\lambda_{\text{stake}}}{n-i}\right)$ and by the tower property for telescoping sums, $\Expect[\stakedist(0)_{(i)}]] = \Theta\left(\frac{\log i}{\lambda_{\text{stake}}}\right)$}, which represents the idea that the $k$th entrant to a network should have a decaying fraction (in this case $\frac{1}{k}$) of the total token supply. 
We note that we do not use power law distributions as there are conflicting reports of the wealth distribution in Bitcoin actually following a power law \cite{kondor2014rich,alvarez2018long}.
Given the statistical difficulty of discerning if one has a power law versus an exponential decay \cite{shalizi2007so,stumpf2012critical} and the extra parameter in a power law (e.g. $p(x = t) \propto x_{\min} t^{-a}$), we decided to use exponential initial distributions.

In our simulations, we kept the initial interest rate at $10\%$, as we saw very little sensitivity to the initial choice.
In particular, agents quickly rebalance their portfolio into lending, if the rate is high enough and this appeared to equilibrate within a small number of time steps.
We swept the other parameters, $\beta_0, \beta_1, \staketime, \lendtime, \slashprob$, over realistic ranges and for each choice of parameter, we sampled trajectories for a variety of seeds $s$ to get an ensemble average. 

\subsection{Simulation Loop}
The main simulation loop follows that as described at the end of \S\ref{sec:statetrans}. 
We exactly evaluate the Markowitz update rule, with constraints, via the exact solution for optimal portfolios \cite{boyd2017multi}.
Given that we are solving $n$ independent, two-dimensional Markowitz problems, we evaluated the constraints analytically (as opposed to using a convex solver like Gurobi or CVX).
The main event loop has the following causal ordering:
\begin{enumerate}
    \item If $t = k \staketime$ for some $k\in\N$, allow agents to update their Markowitz portfolios\footnote{Note that Assumption \ref{as:martingale} allows us to sample the agent portfolio updates in any order}
    \item Sample the borrowing demand distribution
    \item Update $\gamma_t$ as a function of the new borrowing demand
    \item Run Algorithm 1 to determine who wins the reward $R_t$ and/or if they get slashed
\end{enumerate}

\subsection{Borrowing Demand Distributions}\label{sec:bdd}
\begin{figure}
    \centering
    \includegraphics[scale=0.4]{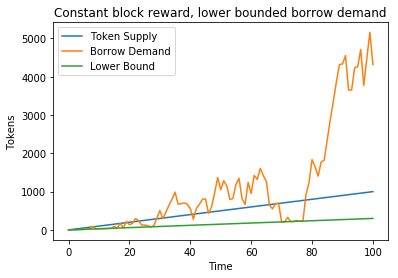}
    \includegraphics[scale=0.4]{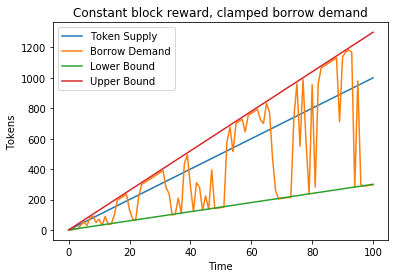}
    \includegraphics[scale=0.4]{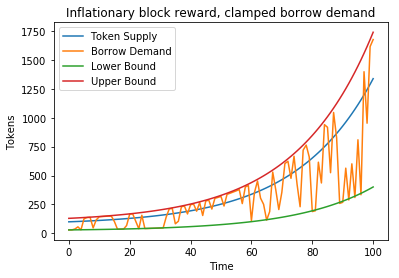}
    \includegraphics[scale=0.4]{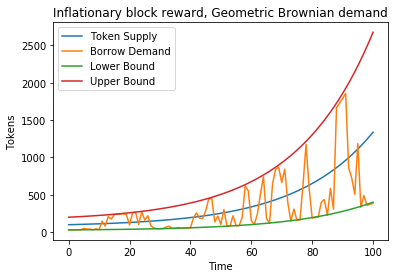}
    \caption{Borrowing demand sample paths, $\xi_t$, juxtaposed with the money supply $S_t$. The sample paths in the upper left (linear, $R_t = c)$, upper right (deflationary, $R_t = kr^t, r < 1$), and lower left (inflationary, $R_t = kr^t, r > 1$) use simple random walks, whereas the figure in the lower right samples a geometric Brownian motion path with reflecting boundary conditions at $(1-\eta_0)S_t, (1+\eta_1)S_t$.}
    \label{fig:bd_sample_paths}
\end{figure}
We sampled a variety of borrowing demand distributions for deflationary and inflationary block rewards, as illustrated via the sample paths in Figure \ref{fig:bd_sample_paths}. 
These stochastic borrowing demand paths each have four parameters: mean, variance, maximum, and minimum demand.
We define the maximum and minimum demand parameters as multipliers of the token supply, so that the minimum is $(1-\eta_0)S_t$ and the maximum is $(1+\eta_1)S_t$ (illustrated via the lower and upper bounds in figure \ref{fig:bd_sample_paths}, respectively).

\subsection{Results}
\begin{figure}
    \begin{tabular}{cccc}
    \includegraphics[scale=0.35]{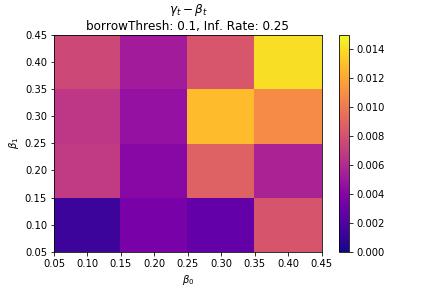} &
    \includegraphics[scale=0.35]{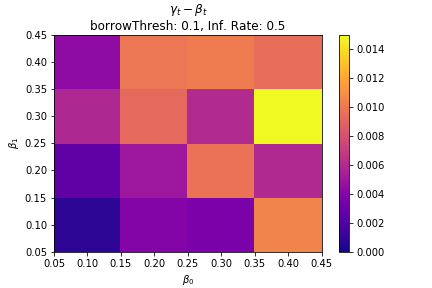} &
    \includegraphics[scale=0.35]{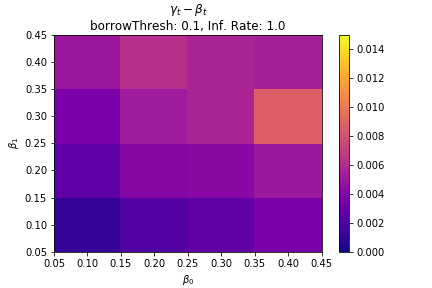} \\
    \includegraphics[scale=0.35]{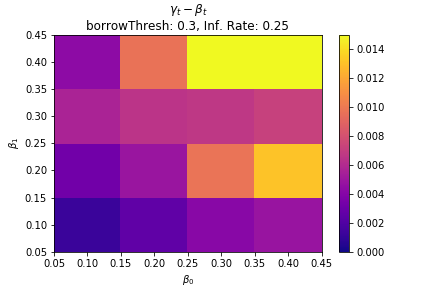} & 
    \includegraphics[scale=0.35]{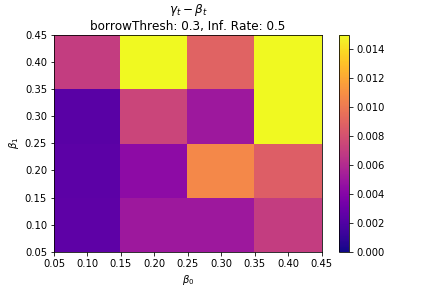} & 
    \includegraphics[scale=0.35]{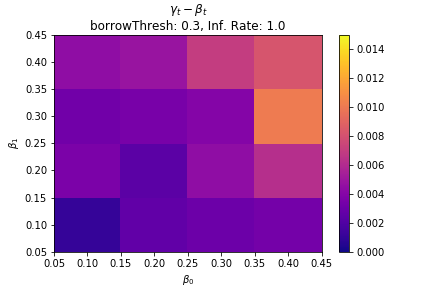} &\\
    \includegraphics[scale=0.35]{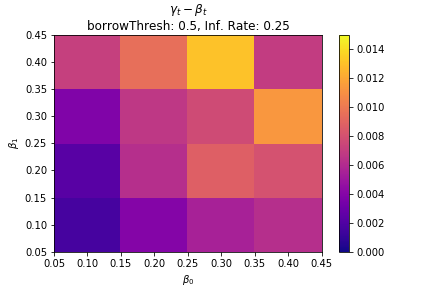} &
    \includegraphics[scale=0.35]{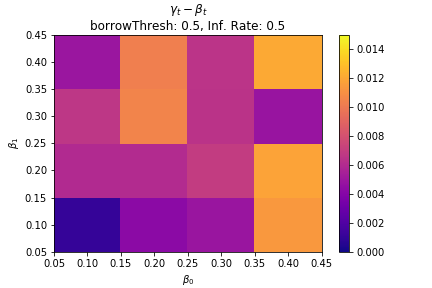} &
    \includegraphics[scale=0.35]{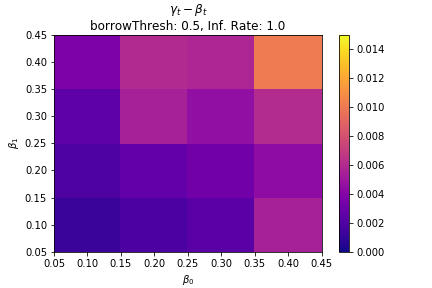} & \\
    \includegraphics[scale=0.35]{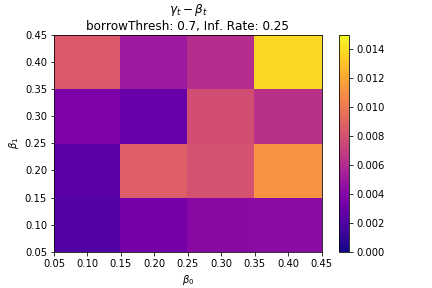} &
    \includegraphics[scale=0.35]{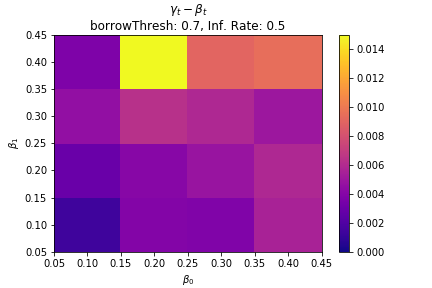} &
    \includegraphics[scale=0.35]{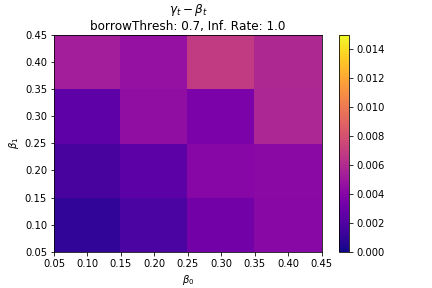}
    \end{tabular}
    \caption{Plots of $g(\beta_0, \beta_1) = \Expect[\gamma_t - \beta_t | \beta_0, \beta_1]$. As you go from left to right, we increase the inflation rate $r$ (recall that when $r < 1$, we are deflationary) and as we go from top to bottom, the borrow threshold increased. Recall that the borrow threshold is the lower bound line from figure \ref{fig:bd_sample_paths}. We see that when the system has linearly increasing block rewards, we receive very tight spreads and a smooth, almost linear $g(\beta_0, \beta_1)$. On the other hand, as we have increasing deflationary policies, we see that we cause the expected spread to become more random and larger.}
    \label{fig:rate_delta}
\end{figure}

\begin{figure}
    \begin{tabular}{cccc}
    \includegraphics[scale=0.35]{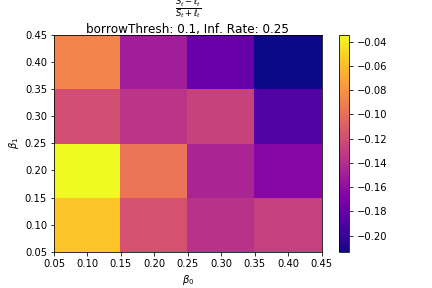} &
    \includegraphics[scale=0.35]{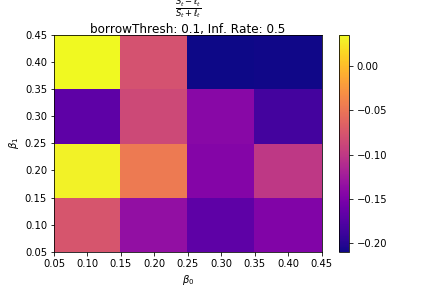} &
    \includegraphics[scale=0.35]{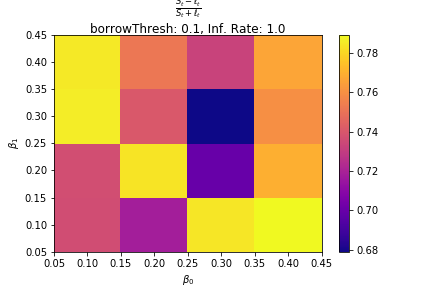} \\
    \includegraphics[scale=0.35]{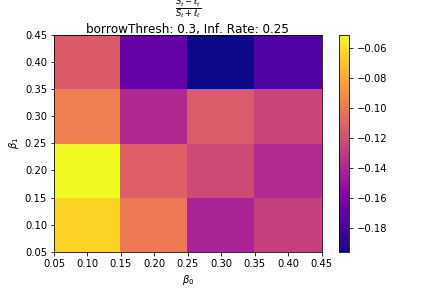} & 
    \includegraphics[scale=0.35]{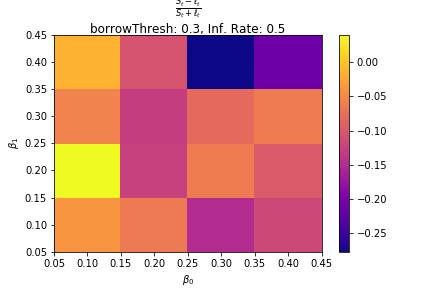} & 
    \includegraphics[scale=0.35]{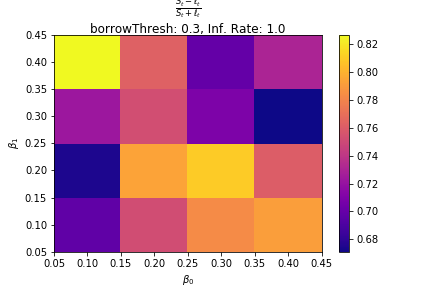} &\\
    \includegraphics[scale=0.35]{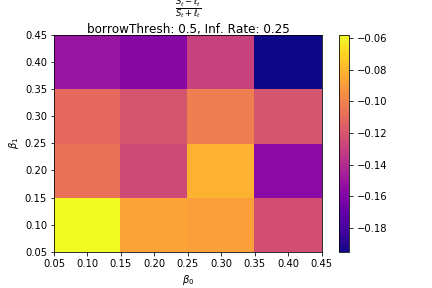} &
    \includegraphics[scale=0.35]{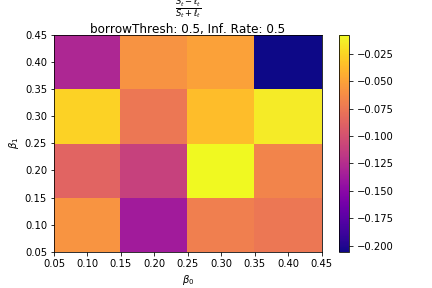} &
    \includegraphics[scale=0.35]{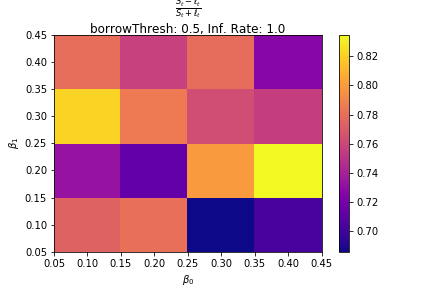} & \\
    \includegraphics[scale=0.35]{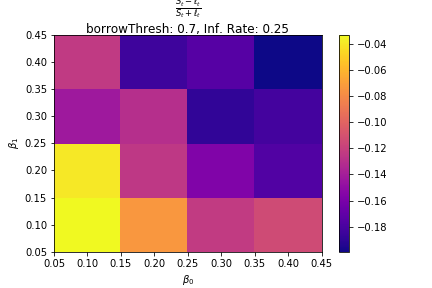} &
    \includegraphics[scale=0.35]{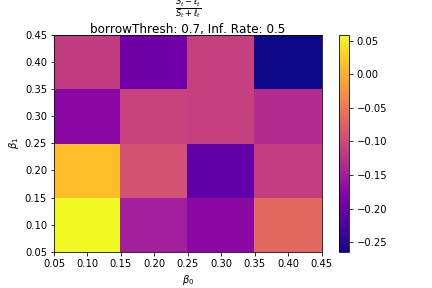} &
    \includegraphics[scale=0.35]{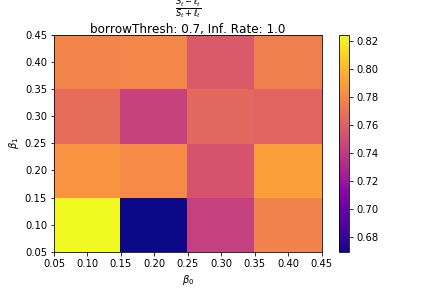}
    \end{tabular}
    \caption{Plots of $f(\beta_0, \beta_1)$ for differing inflation rates and borrowing thresholds. Note that the color bars vary for each figure. We see that when the policy is deflationary, $f(\beta_0, \beta_1) < 0$. Moreover, when the borrowing threshold is high, we cause the plots to become less random and have a more clear dependence on $\beta_0, \beta_1$.}
    \label{fig:dfs}
\end{figure}

In order to evaluate a variety of realistic conditions, we swept through many parameters illustrated above.
The most interesting results come from looking at individual trajectories, such as \ref{fig:lend_cross}, and heatmaps of how certain scalar functions behave as we vary the lending rate parameters $\beta_0, \beta_1$.
We generated heatmaps for number of random seeds and took averages over these random instantiations to generate Figures \ref{fig:rate_delta} and \ref{fig:dfs}.
The two main measurements that we looked at were $f(\beta_0, \beta_1) = \Expect_t\left[\frac{S_t - \ell_t}{S_t+\ell_t} \Bigg| \beta_0, \beta_1, \tau\right]$ and $g(\beta_0, \beta_1) = \Expect_t[\gamma_t - \beta_t | \beta_0, \beta_1, \tau]$, where given a non-anticipating stochastic process $X_t$ and a function stopped at time $\tau$, $\Expect_t[f(X_t) | \tau] = \int_0^{\tau} f(X_t) dX_t$, for the stochastic measure $dX_t$.
We ran all trajectories for $\tau=$250,000 blocks and approximate $\Expect_t[f(X_t) | \tau] \approx \sum_{i=0}^{\tau / \Delta} f(X_i) \xi_i$, where $\xi_i$ is sampled from $dX_{t+\Delta} - dX_t$. 
The first quantity, $f(\beta_0, \beta_1)$, is the normalized difference between the staked supply and the lent supply and when it is greater than zero, there is more quantity staked, on average.
It is normalized relative to the total money supply $S_t + \ell_t$ so that $\forall t, \frac{S_t - \ell_t}{S_t+\ell_t}\in (0,1)$ and thus we can compare the relative staking to lending proportions at different times.
The second quantity, $g(\beta_0, \beta_1)$ measures the linear spread between borrowing and lending rates, which varies over time even though $1-\gamma_0 = \frac{\beta_t-\gamma_t}{\beta_t}$ is constant.
If the market for borrowing is efficient and there is less churn, we should expect that this rate should be quite low.
Note that in all simulations used, we set the relative spread, $1-\gamma_0$, to be 0.5\%.

\begin{figure}
    \centering
    \includegraphics[scale=0.355]{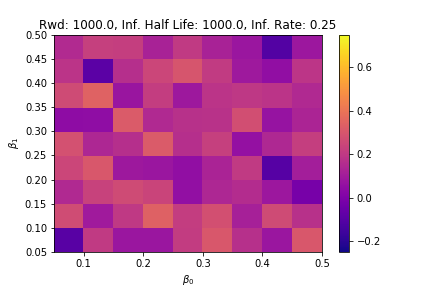}
    \includegraphics[scale=0.355]{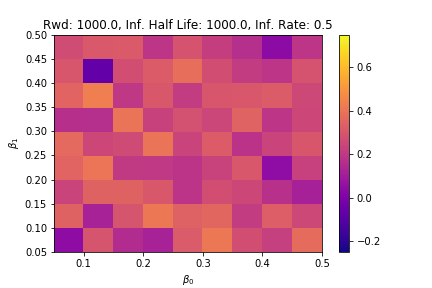}
    \includegraphics[scale=0.355]{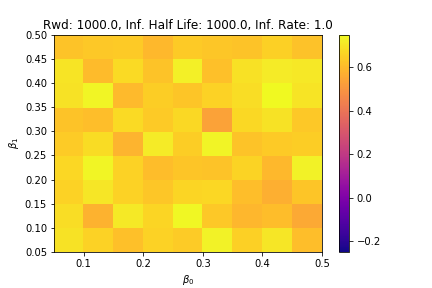}
    \caption{Heat maps of $f(\beta_0, \beta_1) = \Expect_t\left[\frac{S_t - \ell_t}{S_t + \ell_t}\bigg| \beta_0, \beta_1 \right]$ for different deflation schedules at a half-life equal to 0.25\% of the total simulation time (400 Bitcoin-style halvenings occur). Each pixel is the average of 100 random initializations with the other parameters fixed. On the left, we have a very high deflationary rate and we can see that the values are negative, implying that $\ell_t > S_t$, whereas on the far right picture, which is non-deflationary, we see that $S_t > \ell_t$}
    \label{fig:heatmap}
\end{figure}

In Figure \ref{fig:heatmap}, we see plots of $f(\beta_0, \beta_1)$ for different inflation rates $r \in \{0.25, 0.5, 1.0\}$. 
Note that when $r = 1.0$, then we have a linear polynomial inflation rate.
This figure demonstrates that even though the dependence on $\beta_0, \beta_1$ appears to be random, it is clear that the deflationary figures tend to have significantly more lending than staking.
On the other hand, the linearly inflating component enjoys a significant advantage with regards to staked supply.
Note that the dependence on $\beta_0, \beta_1$ appears mostly random because of the use of  a common scale to plot them.
The usage of a common scale dampens the correct scale to plot the figures at (which should be on the order of $\mathsf{Var}(f(\beta_0, \beta_1)$), which varies dramatically as a function of the other parameters that we sample, such as the block reward.
In some of the other plots that we will look at, we use independent color scales for each plot to emphasize the variation as a function of $\beta_0$ and $\beta_1$.

On the other hand, in figure \ref{fig:rate_delta}, we see an array of plots that show how the borrowing spread, $g(\beta_0, \beta_1$ changes as a function of inflation rate and borrowing threshold.
The third column, which is the linearly increasing block rewards regime, demonstrates very tight spreads, suggesting that even changing the borrowing demand threshold causes little variation in spreads.
On the other hand, the deflationary regime is much more sensitive to shocks in borrowing demand, as illustrated by the plots in the upper left hand corner of figure \ref{fig:rate_delta}.
These empirical results validate Claims 2 and 3, as we directly see that lending volatility has a much more muted effect on inflationary systems. 

Finally, in figure \ref{fig:dfs}, we see heatmaps of $f(\beta_0, \beta_1)$ for different borrowing thresholds and inflation rates.
Note that these plots have different scales, unlike \ref{fig:heatmap}, and that the deflationary figures are all mainly negative.
From the first two columns, it becomes clear that as borrowing demand increases, $g(\beta_0, \beta_1)$ becomes increasingly monotonically decreasing in $\beta_0, \beta_1$.
This means that as borrowing demand increases, the system's loss of security to lending increases in a more predictable manner.
On the other hand, the linearly inflating regime does not have this issue and continues to have relatively random and non-monotonic dependence on $\beta_0, \beta_0$ as we adjust the borrowing threshold.

\section{Conclusions and Future Work}
This paper has explored how on-chain lending affects network security in PoS networks by studying how rational multiple period Markowitz optimizing agents optimize their token portfolios.
In particular, we find that `bank runs' can occur when many agents collectively move their tokens from staking contracts to lending contracts even when agents have indpendently drawn risk preferences.
These attacks, which are coordinated only by rational optimization, show that the strictly Byzantine threat model is insufficient to describe security in PoS networks.
Our theoretical and empirical findings show that deflationary PoS tokens are susceptible to attack by non-Byzantine, but rational participants who constantly turnover and rebalance their staking and lending portfolios to chase yield.
Moreover, these findings show that one needs to choose a block reward schedule that increases relative to the yield that these on-chain contracts provide.
Our theoretical results give explicit probabilities for the constant relative borrowing demand regime and show that there is a clear and strong dependence on the lending rate $\gamma_t$ in these probabilities.
We buffet these results with agent-based simulations to confirm that these properties hold in more realistic borrowing scenarios.
In the future, we aim to look at how unbonding times and transaction fees (e.g. gas costs) affect these protocols and how different lending curves will affect staking security.
We also aim to provide more precise theoretical results by relaxing a number of the assumptions that are utilized in this paper.

\section*{Acknowledgments}
The author would like to thank Yi Sun, Rei Chiang, Hasu, Haseeb Qureshi, Ivan Bogatyy, Georgios Konstantopoulos, John Morrow, Tony Salvatore, Wei Wang, Hsien-Tang Kao, Jonathan Reem, Josh Chen, Tina Zhen, Nader Al-Naji, Michael Jordan, Tim Roughgarden, and Matteo Leibowitz for comments and feedback. The author would also like to thank Barnab\'e Monnot and Ariah Klages-Mundt for correcting an earlier error in Claim 1 and for helpful expository comments. Finally, the author would like to acknowledge the helpful input from the reviewers of both the Stanford Blockchain Conference and the MIT Cryptoeconomics Systems journal.

\bibliography{references}
\bibliographystyle{acm} 

\clearpage
\appendix
\section*{Appendix}
\section{Notation}\label{app:notation}
\begin{itemize}
    \item $\Delta^n$ is n-dimensional probability simplex, $\Delta^n = \{ (x_1, \ldots,
            x_n) \in \mathbb{R}^n : \sum_{i=1}^n x_i = 1, \forall i, x_i \geq 0\}$
    \item For any $x \in \mathbb{R}^n$, we define the $p$-norm as $\Vert x \Vert_p = \left(\sum_{i=1}^n |x_i|^p\right)^{1/p}$. 
    \item We turn any nonzero vector $x \in \mathbb{R}^n$ with $x \ge 0$ into a probability distribution by defining $\hat{x} = \frac{x}{\Vert x \Vert_1} \in \Delta^n$.
    \item $S_+^{n} \subset \R^{n\times n}$ is the cone of positive definite, symmetric matrices
    \item $\vee, \wedge$ are the standard join and meet of two elements of a lattice. For example if $a, b \in \R$, $a\vee b = \max(a,b), a\wedge b = \min(a,b)$
    \item We use standard Landau notation \cite{sipser2012introduction} on totally ordered sets $D$: Given functions $f : D \rightarrow \R, g : D \rightarrow \R$, we use the following asymptotic notations:
    \begin{itemize}
        \item $f \in O(g) \iff \exists C > 0, \forall d \in D,\, f(d) \leq C g(d)$
        \item $f \in \Omega(g) \iff \exists c > 0, \forall d \in D,\, f(d) \geq c g(d)$
        \item $f \in o(g) \iff \lim_{d \rightarrow \sup D} \frac{f(d)}{g(d)} = 0$
        \item $f \in \Theta(g) \iff f \in O(g)$ and $f \in \Omega(g)$
    \end{itemize}
\end{itemize}

\section{PoS Algorithm}\label{app:pos}
\begin{algorithm}[H]
  \caption{Epoch-based PoS Simulation}\label{alg:pos_sim}
\begin{algorithmic}
\STATE $\stakedist \leftarrow \stakedist(0) \in \Delta^n$ \hfill  Initial token distribution
\STATE $R \rightarrow \emptyset$ \hfill Initialize current epoch's reward set
\STATE $S \rightarrow \emptyset$ \hfill Initialize current epoch's slash set
\STATE $t \leftarrow 0$
\FOR{$epoch \in \mathbb{N}$} 
\STATE \# Update Staking Distribution
\FOR{$id$ in $\{1, \ldots, n_{stakers}\}$}

\IF{$(id, reward) \in R$}
\STATE $\stakedist[id] += reward$
\ENDIF

\IF{$(id, slash) \in S$}
\STATE $\stakedist[id] -= slash$
\ENDIF

\STATE $R \leftarrow \emptyset, S\leftarrow \emptyset$
 
\STATE
\STATE \# Draw next epoch from stake distribution
\STATE
\STATE $\stakeprob \leftarrow \frac{\stakedist}{\Vert \stakedist \Vert_1}$ \hfill Create normalized probability distribution
\FOR {$block \in \{1, \ldots, \staketime\}$}
\STATE $id \sim \estakedist$ \# Draw a winner from the stake distribution
\STATE $X \sim \mathsf{Bern}(p_{\text{slash}})$
\IF{$X == 1$}
\IF{$\exists (id, s) \in S$}
\STATE $S \leftarrow S \cup \{(id, s - S(id, \stakedist))\}$
\STATE $S \leftarrow S - \{(id, s)\}$
\ELSE 
\IF{$\exists (id, r) \in R$}
\STATE $R \leftarrow R \cup \{(id, r + R(t))\}$
\STATE $R \leftarrow R - \{(id, r)\}$
\ENDIF
\ENDIF
\ENDIF
\STATE t += 1
\ENDFOR
\ENDFOR
\ENDFOR

\end{algorithmic}
\end{algorithm}

\section{Proof of Claim 1}
In Assumption \ref{as:bd_sz}, we assert that for all $t, S_t - \ell_t > \delta S_t$ or $1 - \frac{\ell_t}{S_t} > \delta$ and $\frac{1}{1-\frac{\ell_t}{S_t}} < \frac{1}{\delta}$. Since the function $\frac{1}{1-x}$ is convex on $[0,1]$, we have, for any $x \in [0,1-\delta], s \in [0,1]$, $\frac{1}{1-x} < s \left(\frac{1}{\delta}\right) + (1-s) = s\left(\frac{1-\delta}{\delta}\right) + 1$. Using this bound on the function $\frac{1}{1-\frac{\ell_t}{S_t}}$, we have, for any $s \in [0,1]$:
\[
\frac{1}{S_{t} - \ell_{t}} = \frac{1}{S_{t}} \frac{1}{1-\frac{\ell_{t}}{S_{t}}} < \frac{1}{S_{t}}\left(s\left(\frac{1-\delta}{\delta}\right) + 1\right)
\]
In particular, since $\ell_t < S_t$, we can use this bound at $s = \frac{\ell_{t}}{S_{t}}$ to get
\begin{equation}\label{eq:convexapp}
\frac{1}{S_{t} - \ell_{t}} < \frac{1}{S_{t}}\left(\frac{\ell_{t}}{S_t} \left(\frac{1-\delta}{\delta}\right) + 1 \right)
\end{equation}
This implies that
\[
\mustake(t)_i = \frac{\stakedist(t+1)_i}{S_t - \ell_t} < \frac{\stakedist(t+1)_i}{S_{t}}\left(\frac{\ell_{t}}{S_t} \left(\frac{1-\delta}{\delta}\right) + 1 \right)
\]
Note that the condition $S_t - \ell_t > \delta S_t$ also implies that $\frac{\ell_t}{S_t} < 1-\delta$, so that we can right this as:
\[
\mustake(t)_i = \frac{\stakedist(t+1)_i}{S_t - \ell_t} < \frac{\stakedist(t+1)_i}{S_{t}}\left(\frac{(1-\delta)^2}{\delta} + 1 \right)
\]
On the other hand, since $S_t > S_t - \ell_t$, we also have $\mustake(t)_i > \frac{\stakedist(t+1)_i}{S_{t}}$.
Thus we have
\[
|\mustake(t+1)_i - \mustake(t)_i| = \Theta\left( \frac{\stakedist(t+1)_i}{S_{t+1}} - \frac{\stakedist(t)_i}{S_t}\right)
\]
Let $f(\delta)$ be a constant such that 
\[
|\mustake(t+1)_i - \mustake(t)_i| \leq f(\delta) \left( \frac{\stakedist(t+1)_i}{S_{t+1}} - \frac{\stakedist(t)_i}{S_t}\right)
\]
which exists as per the last statement. Given these equation, we can now bound the single time step rebalance as follows:
\begin{align}
    \Vert \mu_i(t+1) - \mu_i(t)\Vert_1 &= \left| \frac{\stakedist(t+1)_i}{S_t - \ell_t} - \frac{\stakedist(t)_i}{S_t - \ell_t} \right| + |\gamma_{t+1} - \gamma_t| \nonumber \\
    &\leq f(\delta) \left| \frac{\stakedist(t+1)_i}{S_{t+1}} - \frac{\stakedist(t)_i}{S_{t}}  \right| 
    + 
    \beta_1\gamma_0 \left| \left(\frac{\ell_{t+1}}{S_{t+1}}\right)^2 
    - 
    \left(\frac{\ell_{t}}{S_{t}}\right)^2 \right| \nonumber \\
    &= \left| \frac{\stakedist(t+1)_i}{S_t + R_t} 
    - 
    \frac{\stakedist(t)_i}{S_t}  \right| +
    \beta_1\gamma_0 \left| \left(\frac{\ell_{t+1}}{S_t + R_t}\right)^2 
    - 
    \left(\frac{\ell_{t}}{S_{t}}\right)^2 \right| \nonumber \\
    &\leq f(\delta)\left| \frac{\stakedist(t)_i + \dstake(t)}{S_t + R_t} 
    - \frac{\stakedist(t)_i}{S_t}  \right| 
    + \beta_1\gamma_0 
    \left| \left(\frac{\ell_t + \dlend(t)}{S_t + R_t}\right)^2 
    - \left(\frac{\ell_{t}}{S_{t}}\right)^2 \right| \nonumber \\
    &\leq \stakedist(t)_i \left| \frac{f(\delta)}{S_t + R_t} - \frac{f(\delta)}{S_t}\right| + \frac{f(\delta)|\dstake(t)|}{S_t + R_t} + \beta_1\gamma_0 \left| \left(\frac{\ell_t + \dlend(t)}{S_t + R_t}\right)^2 - \left(\frac{\ell_{t}}{S_{t}}\right)^2 \right| \nonumber \\
    &\leq \stakedist(t)_i \frac{kf(\delta)R_t}{(S_t + R_t)^2} + \frac{f(\delta)|\dstake(t)|}{S_t + R_t} \nonumber \\
    &+ \beta_1\gamma_0 \ell_t \left( \frac{\ell_t (2S_t R_t + R_t^2)}{S_t^2 (S_t + R_t)^2} + \frac{1}{(S_t+R_t)^2}\left((2|\dlend(t)| + \frac{\dlend^2(t)}{\ell_t}\right)\right) \nonumber\\
    &\leq \stakedist(t)_i \frac{f(\delta)R_t}{(S_t + R_t)^2} + \frac{f(\delta)|\dstake(t)|}{S_t + R_t} +
    \beta_1\gamma_0\ell_t \left( \Theta\left(\frac{1}{S_t^2}\right) + \frac{2|\dlend(t)|}{(S_t+R_t)^2} + \frac{\dlend^2(t)}{\ell_t} \right) \nonumber \\
    &\leq \frac{f(\delta)}{S_t + R_t}\left(|\dstake(t)| + \frac{2\beta_1\gamma_0\ell_t + R_t}{S_t}\right) + \beta_1\gamma_0 \dlend^2(t) \label{eq:ub}
\end{align}
 
\section{Proof of Claim 2}
Since $R_t = o(2^n), S_t = \sum_{i<t} R_i$, we reach the desired conclusion if for any $\beta \in (0,1], \Prob[|\dstake(t)| > S_t^{\beta_1}] = O(2^{-S_t^\beta})$. Let $W_i(t) = \stakedist(t)_i + \lenddist(t)_i$ be the wealth of agent $i$. Then the unconstrainted Markowitz update gives,
\[
\stakedist(t+1)_i = \frac{\alpha_i \stakedist(t)_i}{S_t - \ell_t} W_i(t)
\]
where $\alpha_i$ is the inverse covariance sample for agent $i$. Thus,
\[
|\dstake(t)| = |\stakedist(t+1)_i - \stakedist(t)_i| = \stakedist(t)_i \left|\frac{\alpha_i W_i(t)}{S_t-\ell_t} - 1\right|
\]
Thus, we have:
\begin{align}
    \Prob\left[|\dstake(t)| > S_t^{\beta}\right] &= \Prob\left[ \stakedist(t)_i \left|\frac{\alpha_i W_i(t)}{S_t-\ell_t} - 1\right| > S_t^{\beta}\right] \\ 
    &= \Prob\left[ \alpha_i > \left(\frac{S_t - \ell_t}{W_i(t)}\right)(S_t^{\beta} + \stakedist(t)_i)\right] < 2^{-\staketime S_t^{\beta}}
\end{align}
where the last step holds since $\alpha_i$ is an exponential random variable with parameter $\staketime$.

\section{Proof of Claim 3}
Recall that $\beta_i$ was defined in equation \ref{eq:cov} and is different than the lending rate $\beta_t$. Note the following, which uses $\beta_i \sim_{\text{iid}} \mathsf{Exp}(\lendtime)$:
\begin{align}
    \Expect[\ell_{t+1}^2 | \mathcal{F}_t] &= \Expect\left[\left(\gamma_t \sum_{i=1}^n \beta_i W_i(t)\right)^2 \bigg| \mathcal{F}_t\right] \\
    &= \gamma_t^2 \Expect\left[\sum_{i=1}^n \beta_i^2 W_i(t)^2 + \sum_{i\neq j}\beta_i \beta_j W_i(t) W_j(t)\right] \\
    &= \gamma_t^2\left( \frac{2}{\lendtime^2} \sum_{i=1}^n W_i(t)^2 + \frac{1}{\lendtime^2}\sum_{i\neq j} W_i(t) W_j(t)\right) \\
    &= \gamma_t^2\left( \frac{1}{\lendtime^2} \left(\sum_{i=1}^n W_i(t)\right)^2 + \frac{1}{\lendtime^2} \sum_{i=1}^n W_i(t)^2\right) \\
    &= \gamma_t^2\left( \frac{S_t^2}{\lendtime^2} + \frac{1}{\lendtime^2} \sum_{i=1}^n W_i(t)^2\right) \\ 
    &= \gamma_t^2\left( \frac{S_t^2}{\lendtime^2} + \frac{\Vert W(t)\Vert_2^2}{\lendtime^2} \right)\label{eq:lbd}
\end{align}
Furthermore, we have the following:
\begin{equation}
    \Expect[\ell_{t+1} | \mathcal{F}_t] = \Expect\left[ \sum_{i=1}^n \gamma_t \beta_i W_i(t) \bigg| \mathcal{F}_t\right] = \frac{\gamma_t S_t}{\lendtime} \label{eq:expect_l} 
\end{equation}
Combining equations \eqref{eq:lbd} and \eqref{eq:expect_l} yields:
\begin{equation}
    \Var[\ell_{t+1}\vert\mathcal{F}_t] = \Expect[\ell_{t+1}^2 \vert \mathcal{F}_t] - \left(\Expect[\ell_{t+1}\vert\mathcal{F}_t]\right)^2 = \frac{\gamma_t^2 \Vert  W(t)\Vert_2^2}{\lendtime^2}
\end{equation}
as claimed.
As the Cauchy-Schwarz inequality implies that $\Vert x \Vert_2^2 \leq \Vert x \Vert_1^2 \leq \sqrt{n} \Vert x \Vert_2^2$ and $S_t = \Vert W(t) \Vert_1$, we have:
\begin{equation}\label{eq:l1_l2}
    1 \leq \frac{S_t^2}{\Vert W(t) \Vert_2^2} = \frac{\Vert W(t) \Vert_1^2}{\Vert W(t) \Vert_2^2} \leq \sqrt{n}
\end{equation}
Thus we have the final part of the claim:
\begin{equation}
    \Var[\ell_{t+1} \vert \mathcal{F}_t] = \frac{\gamma_t \Vert W(t)\Vert_2^2}{\lendtime^2} = \frac{\gamma_t S_t^2}{\lendtime^2}\left( \frac{\Vert W(t)\Vert_2^2}{S_t^2} \right) \in \frac{\gamma_t S_t^2}{\lendtime^2}\left[\frac{1}{\sqrt{n}}, 1\right]
\end{equation}

\section{Proof of Claim 4}
Noting that $\gamma_t = \frac{kS_t}{\ell_t + kS_t} = \frac{1}{1+\frac{\ell_t}{kS_t}}$ implies that for the choice of $k$ in the claim, \eqref{eq:lbd} is greater than $\ell_t^2$ and thus $\ell_t^2$ is a non-negative submartingale with bounded differences (e.g. $\dlend^2(t) \leq S_t^2$).
In particular, note that $\gamma_t \geq \beta_1\left(\frac{k}{k+1}\right)^2 + \beta_0\left(\frac{k}{k+1}\right) = \Omega\left(\frac{k}{k+1}\right)$.
Thus via equation \eqref{eq:lbd}, we have:
\begin{equation}\label{eq:lbd2}
\Expect[\ell_t^2 \vert \mathcal{F}_t] = \frac{\gamma_t^2 S_t^2}{\lendtime^2} \left(1 + \frac{\Vert W(t)\Vert_2^2}{S_t^2}\right) = \frac{\gamma_t^2 S_t^2 \eta_t}{\lendtime^2} \geq C' \left(\frac{k}{k+1}\right)^2 \frac{S_t^2 \eta_t}{\lendtime^2}
\end{equation}
The choice $k = \Omega\left(\frac{\alpha_t}{1-\alpha_t}\right)$ ensures that the right-hand side of equation \eqref{eq:lbd2} is greater than $\ell_t^2$.
The norm equivalence between a non-negative submartingale and its quadaratic variation (e.g. Burkholder's inequality \cite{burkholder1973distribution}) applies to $\ell^2_t$ and implies that $\dlend^2(t)$ is also a submartingale.
Note that this relies on the being able to pass bounds on the quadratic variation process to the individual increments, which only applies to the discrete setting \cite{lowther_2010}.  
The tail bound \eqref{eq:tailb} comes from Doob's submartingale inequality and Kolmogorov's inequality for submartingales \cite{billingsley2008probability}.

\section{Proof of Claim 5}
Recall that $\dlend^2(t)$ is a martingale if $\Expect[\dlend^2(t) \vert \mathcal{F}_t] = \dlend^2(t)$, a supermartingale if $\Expect[\dlend^2(t) \vert \mathcal{F}_t] \leq \dlend^2(t)$, and a submartingale if $\Expect[\dlend^2(t) \vert \mathcal{F}_t] \geq \dlend^2(t)$.
The expansion of $\Expect[\dlend^2(t+1) \vert \mathcal{F}_t]$ is;
\begin{align}\label{eq:increment_exp}
\Expect[\dlend^2(t)\vert\mathcal{F}_t] &= \Expect[(\ell_{t+1} - \ell_t)^2 \vert \mathcal{F}_t] \nonumber \\
&= \Expect[\ell_{t+1}^2 \vert \mathcal{F}_t] - 2\ell_t \Expect[\ell_{t+1} \vert \mathcal{F}_t] + \ell_t^2 \nonumber \\
&= \frac{\gamma_t^2 S_t^2 \eta_t}{\lendtime^2} - 2 \ell_t \frac{\gamma_t S_t}{\lendtime} + \ell_t^2
\end{align}
Define the quadratic function $f : \R \rightarrow \R$ as:
\begin{align}\label{eq:quadratic}
    f(\gamma_t) &= \gamma_t^2 \left(\frac{S_t^2 \eta_t}{\lendtime^2}\right) - \gamma_t \left(\frac{2\ell_t S_t}{\lendtime}\right) + \left(\ell_t^2 - (\ell_t^2 - 2 \ell_t \ell_{t-1} + \ell_{t-1}^2) \right)\nonumber \\
    &=  \gamma_t^2 \left(\frac{S_t^2 \eta_t}{\lendtime^2}\right) - \gamma_t \left(\frac{2\ell_t S_t}{\lendtime}\right) + \left( 2 \ell_t \ell_{t-1} - \ell_{t-1}^2\right)
\end{align}
Note that $f(\gamma_t) = \Expect[\dlend^2(t) \vert \mathcal{F}_t] - \dlend^2(t-1)$ so that the roots of $f$ correspond to when $\dlend^2$ is a martingale.
Moreover, when $f > 0, \dlend^2(t)$ is a submartingale, and a supermartingale otherwise.
Thus, we only need to find the roots of $f$ and to figure out when $f > 0$ and $f < 0$ to classify this transition.
The roots of $f, r_{\pm}$ are:
\[
r_{\pm} = \frac{\ell_t \lendtime}{S_t \eta_t} \left(1 \pm \sqrt{1 + \frac{\eta_t \ell_{t-1}(\ell_{t-1}-2\ell_t)}{\ell_t^2}}\right)
\]
Thus, the roots are real if and only if we have:
\begin{align}
    \eta_t (2\ell_t \ell_{t-1} -\ell_{t-1}^2) < \ell_t^2 \nonumber \\ 
    \iff& \eta_t \ell_{t-1} (\dlend(t-1) + \ell_t) < \ell_t^2 \label{eq:delta_cond} \\
    \iff&  \dlend(t-1) < \frac{\ell_t^2}{\eta_t \ell_{t-1}} - \ell_t \\
    \Rightarrow& \dlend(t-1) < \ell_t \left(\frac{\ell_t}{2\ell_{t-1}} - 1\right)
\end{align}
Recall that for a quadratic function $f(x ) = ax^2 + bx + c$ with real roots $r_{\pm}$, we have $f^{-1}((-\infty, 0)) = [r_-, r_+]$ if $f\left(\frac{r_+ + r_-}{2}\right) = f\left(-\frac{b}{4a}\right) < 0$, otherwise $f^{-1}((-\infty, 0)) = (-\infty, r_-) \cup (r_+, \infty)$.
For equation \eqref{eq:quadratic}, we have $\frac{-b}{4a} = \frac{\ell_t \lendtime}{2 S_t \eta_t}$, which gives:
\begin{align}
    f\left( \frac{\ell_t \lendtime}{2 S_t \eta_t}\right) &= \left(\frac{\ell_t^2 \lendtime^2}{4 S_t^2 \eta_t^2}\right)\left(\frac{S_t^2 \eta_t}{\lendtime^2}\right) - \left(\frac{\ell_t\lendtime}{2 S_t \eta_t}\right) \left(\frac{2 \ell_t S_t}{\lendtime}\right) + \ell_{t-1}(\dlend(t-1) + \ell_t) \nonumber \\
    &= \frac{\ell_t^2}{4\eta_t} - \frac{\ell_t^2}{\eta_t} + \ell_{t-1}(\dlend(t-1) + \ell_t) \nonumber \\
    &= -\frac{3 \ell_t^2}{4\eta_t} + \ell_{t-1}(\dlend(t-1) + \ell_t) \label{eq:midpt}
\end{align}
Note that equation \eqref{eq:midpt} is less than zero when
\[
\dlend(t-1) < \ell_t \left(\frac{3\ell_t}{4\ell_{t-1}} -1\right)
\]
which is already subsumed by the condition for having real roots.
Thus, provided that $r_{\pm}\in (0,1)$, we have the desired conclusion. 

\section{Proof of Claim 6}
As before, let the $i$th agent's token wealth be denoted $W_i(t) = \stakedist(t)_i + \lenddist(t)_i$. Using the Markowitz update rule, we have the following:
\begin{align}
    \mustake(t)_i &= 
    \frac{\stakedist(t+1)_i}{S_{t+1} - \ell_{t+1}} 
    - 
    \frac{\stakedist(t)_i}{S_t - \ell_t} 
    = 
    \frac{\stakedist(t)_i}{S_t - \ell _t}
    \frac{\alpha_i W_i(t)}{S_{t+1}-\ell_{t+1}} 
    - 
    \frac{\stakedist(t)_i}{S_t-\ell_t} 
    \nonumber\\ 
    &= 
    \frac{\stakedist(t)_i}{S_t-\ell_t}
    \left(\frac{\alpha_i W_i(t)}{S_{t+1}-\ell_{t+1}} -1 \right) 
    \label{eq:diff} 
\end{align}
Applying \eqref{eq:convexapp} at $s = \frac{\ell_{t+1}}{S_{t+1}}$ to \eqref{eq:diff} gives:
\begin{align}
\mustake(t)_i = 
\frac{\stakedist(t)_i}{S_t-\ell_t}
\left(\frac{\alpha_i W_i(t)}{S_{t+1}-\ell_{t+1}} -1 \right) 
&= 
\frac{\stakedist(t)_i}{S_t-\ell_t}
\left( \frac{\alpha_i W_i(t)}{S_{t+1}} \frac{1}{1-\frac{\ell_{t+1}}{S_{t+1}}} - 1 \right) 
\nonumber \\ 
&< \frac{\stakedist(t)_i}{S_t-\ell_t}
\left( \frac{\alpha_i W_i(t)}{S_{t+1}} 
\left(\frac{\ell_{t+1}}{S_{t+1}} 
    \left(\frac{1-\delta}{\delta}
    \right) + 1
    \right) - 1 
\right)
\end{align}
Recall that $\ell_{t+1} = \gamma_t \sum_{i=1}^{n} \beta_i W_i(t)$ and that for positive functions $f \geq g$, $\Prob[f < \epsilon] \leq \Prob[g < \epsilon]$.
Note that $\stakedist(t)_i$ as adapted to a natural filtration $\mathcal{F}_t$ that is non-anticipating, by construction\footnote{As we sample the PoS algorithm and Markowitz weights randomly (and not psuedorandomly), there is no advantage to looking into the future.}.
This means that we can compute probabilities conditional on $\mathcal{F}_t$, e.g. $\Prob[\ell_t | \mathcal{F}_t] = \ell_t$, 
Using this and recalling that the money supply $S_t$ is a deterministic function yields:
\begin{align}
    \Prob[|\mustake(t)_i| < \epsilon | \mathcal{F}_t] 
    &\geq \Prob\left[ 
        \left|\frac{\stakedist(t)_i}{S_t-\ell_t}
            \left( \frac{\alpha_i W_i(t)}{S_{t+1}} 
                \left(\frac{\ell_{t+1}}{S_{t+1}} \left(\frac{1-\delta}{\delta}\right)
                + 1\right)
                - 1 \right)\right| < \epsilon \Bigg| \mathcal{F}_t\right] \label{eq:initprob}\\
    &\geq \Prob\left[ 
        \left|\frac{\stakedist(t)_i}{S_t-\ell_t}
            \left( \frac{\alpha_i W_i(t)}{S_{t+1}}
                \left(\frac{\ell_{t+1}}{S_{t+1}}
                    \left(\frac{1-\delta}{\delta}
                    \right)
                \right)
            \right)
        \right| < \epsilon \Bigg| \mathcal{F}_t
    \right](1-e^{-\frac{S_{t+1}}{W_i(t)}}) \label{eq:second}\\
    &\geq \left(1-e^{-S_{t+1} / W_i(t)}\right)\Prob\left[
        \alpha_i \sum_{i=1}^n \beta_i  < \epsilon 
        \left(\frac{S_t - \ell_t}{P_i(t)}\right)
        \left(\frac{S_{t+1}}{\gamma_t}\right)
        \left(\frac{\delta}{1-\delta}\right)
        \left(\frac{1}{W_i(t) (\max_i W_i(t))}\right)
        \Bigg| \mathcal{F}_t
        \right] \label{eq:third}
\end{align}
where \eqref{eq:second} comes from splitting equation \eqref{eq:initprob} into two terms by conditioning on $A = \left\{\frac{\alpha_i W_i(t)}{S_{t+1}} - 1 < 0\right\}$ and $A^c$ and throwing away the term dependent on $A^c$ and the third inequality \eqref{eq:third} comes from $W_i(t) < S_{t+1}$ and majorizing the sum $\sum_i \beta_i W_i(t)$. 
Note that \eqref{eq:third} is the CDF for the product of an exponentially distributed random variable $\alpha$ and an Erlang distributed random variable, $\sum_{i=1}^n \beta_i$. Let $A \sim \mathsf{Exp}(\staketime)$ and $B \sim \mathsf{Erlang}(n, \lendtime)$. 
Then the product CDF\footnote{Recall that the density of an exponential random variable with parameter $\alpha$ is $\alpha e^{-\alpha x}$ and the PDF of an Erlang distribution with parameters $n, \lambda$ is $\frac{\lambda^n}{(n-1)!} x^{n-1} e^{-\lambda x}$} is:
\begin{align}
    \Prob[AB \leq y] &= \int_0^{\infty} \frac{\staketime^n}{(n-1)!} x^{n-1} e^{-\lendtime x} \Prob\left[A < \frac{y}{x}\right] dx \nonumber \\
                     &= 1 - \frac{\staketime^n}{(n-1)!}\int_0^{\infty} x^{n-1} e^{-\staketime x - \lendtime \frac{y}{x}} dx \nonumber \\
                     &= 1 - \frac{\staketime^n}{(n-1)!}\int_0^{\infty} e^{-\staketime x - \lendtime \frac{y}{x} - (n-1) \log x} dx \label{eq:max}
\end{align}
The integrand in \eqref{eq:max} is maximized around the saddle point of the large deviation function $\Lambda(x) = \staketime x - \frac{\lendtime y}{x} - (n-1)\log x$. Let $x^* = \argmax \Lambda(x)$. Using the first-order condition $\nabla \Lambda(x) = 0$ and the quadratic formula, we get:
\[
x^* = \frac{n-1}{2 \lendtime} \left(\sqrt{1 + \frac{4\staketime\lendtime y}{(n-1)^2}} - 1\right)
\]
From Assumption \ref{as:num_agents}, we know that $4 \staketime \lendtime < (n-1)^2$, so $x^* \approx 2\staketime y$. Applying the method of steepest descent \cite{fedoryuk1987asymptotic} to \eqref{eq:max} implies that there exists $C > 0$ such that $\int_0^{\infty} e^{-\staketime x - \lendtime \frac{y}{x} - (n-1) \log x} dx  < C e^{-\Lambda(x^*)}$ so that we have:
\begin{equation}\label{eq:erlang_bound}
\Prob[AB \leq y] \geq 1 - \frac{C \staketime^n}{(n-1)!} e^{-\Lambda(X^*)} = 1 - \frac{C \staketime^n}{(n-1)!} (2\staketime y)^{n-1} e^{-\staketime^2 y + \frac{\lendtime}{2\staketime}}
\end{equation}
Combining \eqref{eq:third} with \eqref{eq:erlang_bound} gives the result.

\end{document}